\documentclass[preprint,12pt,3p]{elsarticle}
\usepackage{amssymb}
\usepackage{amsmath}
\usepackage{subfigure}
\usepackage{float}
\usepackage{multirow}

\usepackage[]{siunitx}
\usepackage{caption}
\DeclareMathOperator{\sign}{sign}
\graphicspath {{Figures/}}

\journal{}
\begin{document}

\begin{frontmatter}

\title{Anticipation in a velocity-based model for pedestrian dynamics}

\author[label1]{Qiancheng Xu\corref{cor1}}
\address[label1]{Institute for Advanced Simulation,\\ Forschungszentrum J\"ulich, 52425 J\"ulich, Germany}

\cortext[cor1]{corresponding author}

\ead{q.xu@fz-juelich.de}

\author[label1]{Mohcine Chraibi}
\ead{m.chraibi@fz-juelich.de}

\author[label1,label2]{Armin Seyfried}
\address[label2]{School of Architecture and Civil Engineering,\\University of Wuppertal, 42119 Wuppertal, Germany}
\ead{a.seyfried@fz-juelich.de}

\begin{abstract}

Lane formation in bidirectional pedestrian streams is based on a stimulus-response mechanism and strategies of navigation in a fast-changing environment. Although microscopic models that only guarantee volume exclusion can qualitatively reproduce this phenomenon, they are not sufficient for a quantitative description. To quantitatively describe this phenomenon, a minimal anticipatory collision-free velocity model is introduced. Compared to the original velocity model, the new model reduces the occurrence of gridlocks and reproduces the movement of pedestrians more realistically. For a quantitative description of the phenomenon, the definition of an order parameter is used to describe the formation of lanes at transient states and to show that the proposed model compares relatively well with experimental data. Furthermore, the model is validated by the experimental fundamental diagrams of bidirectional flows.
\end{abstract}

\begin{keyword}
anticipation, velocity-based model, pedestrian dynamics, bidirectional flow, lane formation
\end{keyword}

\end{frontmatter}

\section{Introduction}
In bidirectional flow situations, pedestrians self-organize into dynamically varying and separated lanes~\cite{yamori1998going, helbing2001self, hoogendoorn2005self, feliciani2018universal, boltes2018empirical, adrian2019glossary}.
Although the mechanisms behind this apparently organized separation of the crowd are not known for certain and in many cases may seem random, we observe that this formation leads to a reduction in collisions and thus increases the speed.
Unlike with car traffic, where stable lanes are predetermined by the restrictions established by the infrastructure, in pedestrian dynamics, lanes are formed dynamically and naturally with neither external synchronization nor any prior agreement between pedestrians.

We also see lane formation in systems of inanimate particles~\cite{dzubiella2002lane,leunissen2005ionic,sutterlin2009dynamics}, where models ensuring volume exclusion are sufficient to reproduce the phenomenon.
Therefore, with a simple social force model considering the repulsion between particles, Helbing et al.~\cite{helbing1995social} qualitatively reproduced the lane formation in a corridor with periodic boundary conditions.
They attributed two reasons to this phenomenon, the sideways movement, which separates agents moving in opposite directions, and the weak interaction between agents moving in the same lane, which maintains the lanes that have been formed~\cite{helbing2001self}.
However, pedestrians usually avoid collisions by using a stimulus-response mechanism to anticipate changes in the environment. 
Therefore, it is clear that a model based solely on volume exclusion is oversimplified and not suitable for quantitatively reproducing the lane formation in pedestrian systems.

Pedestrians anticipate the changes in their environment by predicting the movement of neighboring pedestrians and they take this information into account in their steering to avoid imminent collisions. 
The effect of anticipation on the movement of pedestrians in bidirectional flows has been discussed and addressed in several works to date.
For instance, Suma et al.~\cite{suma2012anticipation} conducted a bidirectional flow experiment where participants are asked to use cell phones (weak anticipation) or move cautiously (excessive anticipation), to study how anticipation affects the movement of pedestrians.
They found anticipation significantly affects the time it takes for pedestrians to pass through the corridor, and there is an optimal degree of anticipation to realize the minimum passing time.
However, since the scale of the experiment was small, the lane formation was not analyzed quantitatively.
Consequently, the authors proposed an anticipation floor field cellular automata model, which considers the area occupied by agents in the future, to give a more realistic picture of the behavior of agents in the bidirectional flow simulation.
The model was further analyzed in~\cite{nowak2012quantitative} by using an order parameter, which is originally used to detect lanes in a colloidal suspension~\cite{rex2007lane}. 
A quantitative analysis showed that the model with anticipation can reproduce lane formation in higher density situations than the model without anticipation. 
Bailo et al.~\cite{bailo2018pedestrian} also proposed a microscopic model with anticipation, which takes into consideration the time to collision between agents. 
Murakami et al.~\cite{murakami2019levy} performed the bidirectional experiment in a corridor with open boundary conditions, and they observed that the sideways movement of pedestrians before lane formation can be described in terms of the L{\'e}vy walk process.
Therefore, the authors suggested that this sideways movement is strongly related to lane formation.
Moreover, they assumed the most likely action underlying the sideways movement is anticipation.
Then, the relationship between anticipation and lane formation is further studied in~\cite{Murakamieabe7758} through a larger scale bidirectional flow experiment, where pedestrians distracted by cell phones are located at different positions to represent situations with different degrees of anticipation.
They found anticipation favors the formation of lanes in bidirectional flow situations.

An additional strategy to avoid collisions is preferring to follow pedestrians moving in the same direction. 
This strategy is part of the anticipation process to reduce collisions but it reduces the probability of conflicts in a larger time scale than the strategy that only avoids collisions with neighboring pedestrians.
A model that supports this statement was published by Isobe et al.~\cite{isobe2004experiment} who using a bidirectional flow experiment with open boundary conditions, measured the mean time for pedestrians to pass through the corridor and the mean speed of pedestrians during the process.
The authors proposed a lattice gas model taking into account the strategy of following others.
The front area of an agent is divided into three parts and the agent tends to move into the area that contains the highest number of agents moving in the same direction.  
This model reproduces lane formation and the measured experimental data very well.

In addition, the effect of various factors such as flow ratio and heterogeneity of agents on lane formation were studied in several works.
Mossaid et al.~\cite{moussaid2012traffic} performed a bidirectional flow experiment in a ring-shaped corridor to realize periodic boundary conditions.
The authors discovered the dynamic and transient nature of lane formation is due to the heterogeneity of pedestrians' walking speed, and the lane is stable when all pedestrians move at the same speed.
This conclusion was validated using computer simulations.
Feliciani et al.~\cite{feliciani2016empirical} conducted a bidirectional flow experiment in a corridor with open boundary conditions, where various flow ratios of pedestrians in two directions were adopted.
Their result shows that, once lanes are formed, a balanced bidirectional flow, i.e., when the number of pedestrians moving in both directions is equal, is the most stable situation.  

Another phenomenon related to bidirectional flow is the jamming transition, also called gridlock, appearing at a critical density. 
Muramatsu et al.~\cite{muramatsu1999jamming} used a lattice gas model without backstepping to study the jamming transition in bidirectional pedestrian flow with open boundary conditions.
They found the jamming transition does not depend on the corridor size but it is affected by the strength of the drift (the preference to move in the desired direction) and the traffic rule adopted (such as keep to the right).
Fang et al.~\cite{weifeng2003simulation} adopted a cellular automata model with backstepping and the right-hand side rule.
They observed the critical density of jamming transition increases with a higher probability of backstepping.
Nowak et al.~\cite{nowak2012quantitative} studied the phenomenon with the anticipation floor field cellular automata model proposed in~\cite{suma2012anticipation}.
They discovered the anticipation mechanism in the model suppresses the formation of jamming (facilitating the formation of lanes), which leads to an increase in the critical density of the jamming transition.  
However, the jamming transition is only observed in computer simulations.

Furthermore, the fundamental diagram is used to analyze bidirectional streams.
In some early studies summarized in~\cite{zhang2012ordering}, it is believed that there is no clear or only a small difference between uni- and bidirectional flows.
Helbing et al.~\cite{helbing2005self} concluded that bidirectional flows are more efficient than unidirectional flows.
The possible reason behind this is better coordination between people in bidirectional situations (lane formation).
Kretz et al.~\cite{kretz2006experimental} also found that pedestrians use space more efficiently in bidirectional situations.
Subsequently, Zhang et al.~\cite{zhang2012ordering} carried out both uni- and bidirectional flow experiments under laboratory conditions.
A clear difference between the fundamental diagrams of uni- and bidirectional flows is observed when the density is higher than $\SI{1.0}{\per\square\meter}$.
The specific flow reaches a peak with increasing density in the unidirectional flow, whereas a plateau is formed in the bidirectional flow.
However, there are no experimental data for densities higher than $\SI{4.5}{\per\square\meter}$.

In order to reproduce bidirectional flow quantitatively, the anticipation velocity model (AVM) for pedestrian dynamics is proposed.
The action anticipating changes of neighboring pedestrians' positions and the strategy of following others are covered in this model.
The new model is compared to two similar models from the literature, the collision-free speed model and generalized collision-free velocity model, and we highlight the reasons behind the difference.
Moreover, we use the AVM to study the jamming transition, lane formation, and fundamental diagrams in bidirectional flow scenarios.
In the following section, the AVM is described. 

\section{Definition of the anticipation velocity model}
\label{sec:modelDef}
In this model, an agent is represented as a disk with a constant radius $r$.
The position and velocity of pedestrian $i$ are denoted by $\vec{x}_i$ and $\vec{v}_i$, respectively, where $\vec{v}_i=\dot{\vec{x}}_i$.
Furthermore, $\vec{v}_i=\vec{e}_i\cdot v_i$, where $\vec{e}_i$ and $v_i$ denote the direction of movement and the speed of agent $i$, respectively.
Both variables are modeled differently as explained in the following subsections.

\subsection{Submodel for operational navigation}
\begin{figure}[H]
    \centering
    \includegraphics[width=0.8\linewidth]{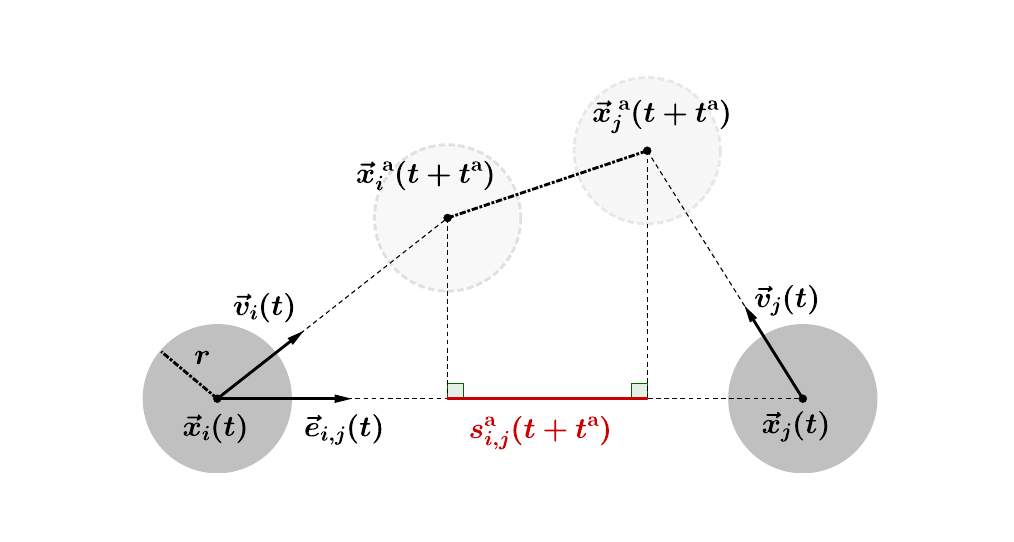}
    \caption{
    An example of ${s}^\text{a}_{i,j}(t+t^\text{a})$, the predicted distance between agents $i$ ($\vec{x}_i$, $\vec{v}_i$) and $j$ ($\vec{x}_j$, $\vec{v}_j$). 
    When pedestrians move towards each other their predicted distance is smaller than the actual distance.
    }
    \label{fig:antiInStrengh}
\end{figure}

The direction of movement of agent $i$ is determined by its desired direction which is a unit vector denoted by $\vec{e}_i^{~0}$ pointing towards its target.
The determination of the target follows various tactical strategies, which is not the subject of the present study.
For operational navigation to avoid collisions and obstructions, in the presence of other agents, the direction of $i$ will deviate from its desired direction $\vec{e}_i^{~0}$. 
To consider anticipation, the process can be divided into the following parts: a. perception of the actual situation, b. prediction of a future situation, and c. selection of a strategy leading to an action.

a. Perception of the actual situation: 
To consider restrictions using visual perception, it is assumed that only agents located in the union of two half-planes, where $i$ is moving or intends to move, affect its direction.
The set containing all agents who have an impact on $i$'s direction of movement is
\begin{equation}
    N_i(t)=\bigg\{j,~\vec{e}_i(t) \cdot \vec{e}_{i,j}(t)>0\ \text{or}\  \vec{e}_i^{\ 0}(t) \cdot \vec{e}_{i,j}(t)>0\bigg\},
\end{equation}
where $\vec{e}_{i,j}$ denotes the unit vector from $i$ to $j$.

b. Prediction of a future situation: 
To consider the prediction, it is assumed that the strength of $j$'s impact on $i$ is a function of the predicted distance between these two agents at a particular time point. 
Given a time constant $t^\text{a}$, which can be interpreted as the prediction time, the predicted distance is defined as 
\begin{equation}
    {s}^\text{a}_{i,j}(t+t^\text{a})=\max \bigg\{2r,~\Big(\vec{x}^\text{~a}_j(t+t^\text{a})-\vec{x}^\text{~a}_i(t+t^\text{a})\Big) \cdot \vec{e}_{i,j}(t) \bigg\},
    \label{equ:sdelta}
\end{equation}
where $\vec{x}^\text{~a}_i(t+t^\text{a})=\vec{x}_i(t)+\vec{v}_i(t)\cdot t^\text{a}$.
See Figure~\ref{fig:antiInStrengh}.

c. Selection of a strategy leading to an action: 
After the introduction of the predicted distance in Eq.~\ref{equ:sdelta}, 
the strength of the impact from agent $j$ on the direction of movement of agent $i$ is defined as
\begin{equation}
\label{equ:R}
    R_{i,j}(t)= \alpha_{i,j}(t) \cdot \exp 
    \bigg( \frac{2 r-s^\text{a}_{i,j}(t+t^\text{a})}{D} \bigg),
\end{equation}
where $D>0$ is a constant parameter used to calibrate the range of the impact from neighbors and $\alpha_{i,j}$ is a directional dependency used to vary the strength of impact from different neighbors (see Eq.~\ref{equ:alpha}).
\begin{equation}
\label{equ:alpha}
    \alpha_{i,j}(t)=k \Big(1+ \frac{1- \vec{e}_i^{~0}(t) \cdot \vec{e}_j(t)}{2}\Big),\; k>0 ,
\end{equation}
where $\alpha_{i,j}$ is minimal ($k$) when both vectors $\vec{e}_i^{~0}$ and $\vec{e}_j$ are aligned and is maximum ($2 k$) when they are anti-aligned, which means that agents influence each other's direction strongly in bidirectional scenarios.
Here, $\alpha_{i,j}$ means agents have a high tendency to follow the agents who move in the same direction. 
When this strategy is used, the probability of further conflicts is reduced. 

\begin{figure}[H]
    \centering
    \includegraphics[width=0.5\linewidth]{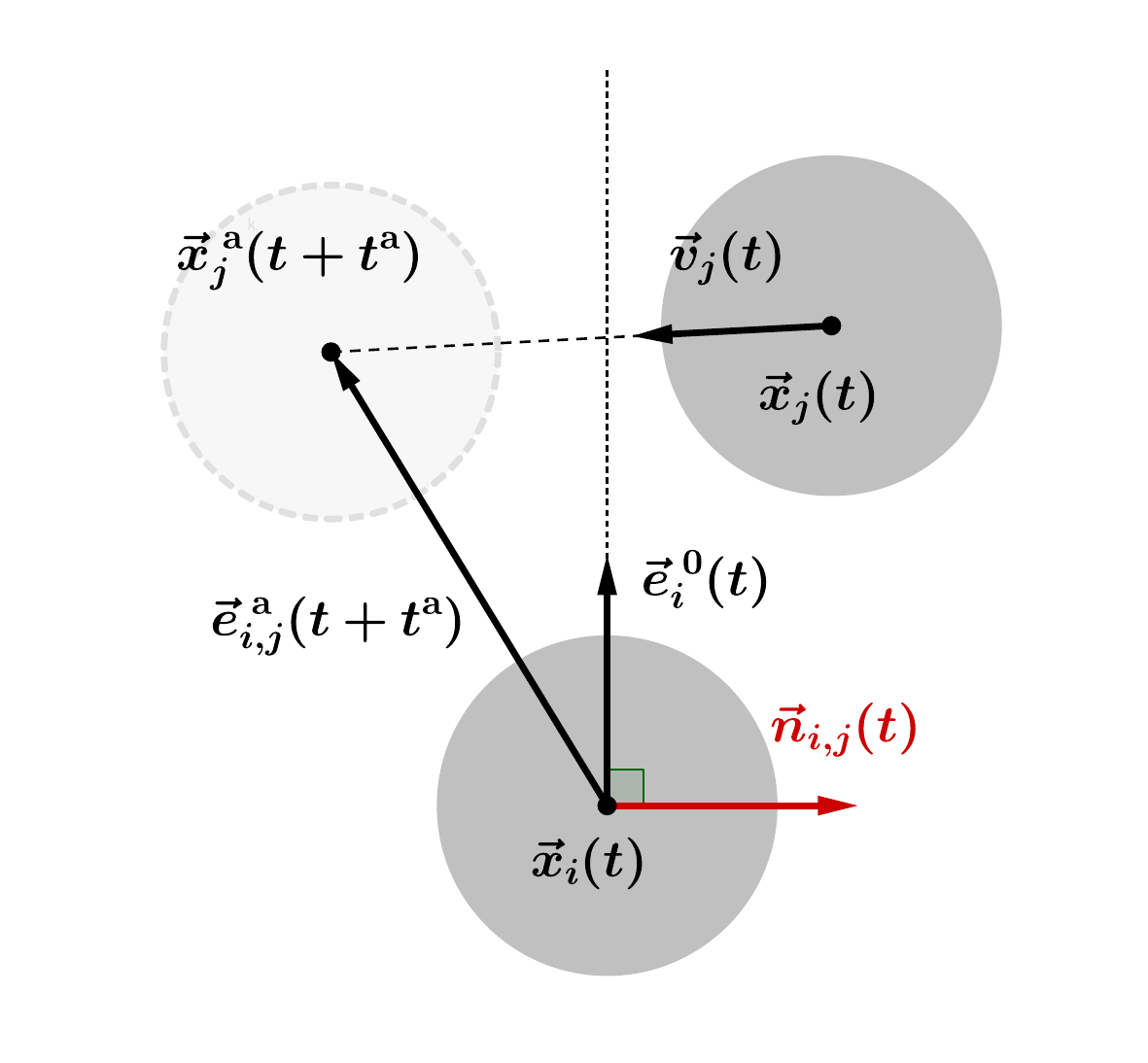}
    \caption{
    The direction of the impact from $j$ on the direction of movement of $i$ ($\vec{n}_{i,j}$) according to Eq.~\ref{equ:nji}.
    }
    \label{fig:antiInDirection}
\end{figure}

The direction of the impact from agent $j$ on $i$'s direction of the movement is defined as
\begin{equation}
\label{equ:nji}
 \vec{n}_{i,j}(t) =-\sign\bigg(\vec{e}^\text{~a}_{i,j}(t+t^\text{a}) \cdot \vec{e}_i^{~0\bot}(t)\bigg) \cdot\vec{e}_i^{~0\bot}(t),
\end{equation}
where $\vec{e}^\text{~a}_{i,j}(t+t^\text{a})= \vec{x}^\text{~a}_j(t+t^\text{a})- \vec{x}_i(t)$.
The direction of $\vec{n}_{i,j}$ depends on the predicted position of agent $j$ after a period of time $t^\text{a}$.
Note that when this predicted position is aligned with the desired direction of $i$, the direction of $\vec{n}_{i,j}(t)$ in Eq.~\ref{equ:nji} is chosen randomly as $\vec{e}_i^{~0\bot}$ or $-\vec{e}_i^{~0\bot}$.
See Figure~\ref{fig:antiInDirection}.
This rule prevents agents from moving in the opposite direction to the desired direction.

Finally, Eq.~\ref{equ:R} and Eq.~\ref{equ:nji} yield the optimal direction of agent $i$ as
\begin{equation}
\label{equ:optimalD}
    \vec{e}_i^\text{~d}(t)=u \bigg(\vec{e}_i^{~0}(t)+\sum_{j\in N_i(t)} R_{j,i}(t)\cdot \vec{n}_{j,i}(t)\bigg),
\end{equation}
where $u$ is a normalization constant such that $\lVert\vec{e}_i^\text{~d}\rVert=1$.
Then, the direction of movement of agent $i$ is updated as
\begin{equation}
\label{equ:direction}
    \frac{d\vec{e}_i(t)}{dt} = \frac{ \vec{e}_i^\text{~d}(t)-\vec{e}_{i}(t)}{\tau},
\end{equation}
where $\tau$ is a relaxation parameter adjusting the rate of the turning process from the current direction $\vec{e}_i$ to the optimal direction $\vec{e}_i^{~d}$.

\subsection{Submodel for the speed}
After obtaining the new direction of the movement according to Eq.~\ref{equ:direction}, the set of neighbors that are imminently colliding with $i$ is defined as
\begin{equation}
    J_i=\Big\{j,~\vec{e}_i \cdot \vec{e}_{i,j} \ge 0\ \text{and}\ \left|\vec{e}_i^{~\bot} \cdot \vec{e}_{i,j}\right| \leq \frac{2 r}{s_{i,j}}\Big\},
\end{equation}
where $s_{i,j}$ is the current distance between $i$ and $j$.
Therefore, the maximum distance that agent $i$ can move in the direction without overlapping other agents is
\begin{equation}
\label{equ:spacing in front i}
    s_i=\min_{j\in J_i}s_{i,j}-2r.
\end{equation}

Finally, the speed of agent $i$ in the new direction is
\begin{equation}
\label{equ:vf}
    v_i=\min\Big\{v_i^0,~\max\big\{0,\frac{s_{i}}{T}\big\}\Big\},
\end{equation}
where $v_i^0$ is the free speed of agent $i$, and $T>0$ is the slope of the speed-headway relationship.
The speed submodel used here is the same as in the generalized collision-free velocity model.

\section{Test with two interacting agents}
\label{sec:binary}
Binary interaction scenarios with the collision-free speed model (CSM), the generalized collision-free velocity model (GCVM), and the AVM, respectively, are studied to assess the models' ability.
The three models adopt the same speed submodel but different submodels for operational navigation.
In both the CSM and the GCVM, the strength of the effect from agent $j$ on agent $i$'s direction of movement is a function of the distance between the two agents.
As for the direction of this effect, in the CSM it is from $j$ to $i$, while in the GCVM it is obtained with Eq.~\ref{equ:nji} ($t_a=\SI{0}{\second}$).
A detailed introduction to the CSM can be found in~\cite{tordeux2016collision} and to the GCVM in~\cite{xu2019generalized}. 
The parameters of these models are summarized in Table~\ref{tab:parameterCompare}.

\begin{table}[H]
    \centering
    \begin{tabular}{|c|c|c|c|c|c|c|c|}
    \hline
          & $r$ [\SI{}{\meter}] & $k$ Eq.~\ref{equ:alpha} & $D$ [\SI{}{\meter}] Eq.~\ref{equ:R} & $T$ [\SI{}{\second}] Eq.\ref{equ:vf} & $\Delta t$ [\SI{}{\second}] & $\tau$ [\SI{}{\second}] Eq.~\ref{equ:direction} & $t^{\text{a}}$ [\SI{}{\second}] \\
   \hline
         CSM & \multirow{3}{*}{0.18} & \multirow{3}{*}{3} & \multirow{3}{*}{0.1} & \multirow{3}{*}{1.06} & \multirow{3}{*}{0.05} & $\setminus$ & $\setminus$ \\
    \cline{1-1}\cline{7-8}
        GCVM & & & & & & \multirow{2}{*}{0.3} & $\setminus$ \\
    \cline{1-1}\cline{8-8}
         AVM & & & & & & & 1 \\
    \hline
    \end{tabular}
    \caption{
    The parameters of the models in binary interaction simulations.
    Here, $r$ is the radius of agents, $\Delta t$ is the time step size, and $t_a$ is the prediction time. 
    The simulation in the present study is conducted using the Euler scheme.
    Parameter values of the CSM and the GCVM are obtained from~\cite{xu2019generalized}.
    }
    \label{tab:parameterCompare}
\end{table}

\begin{figure}[H]
    \centering
    \subfigure[]{\includegraphics[width=0.5\linewidth]{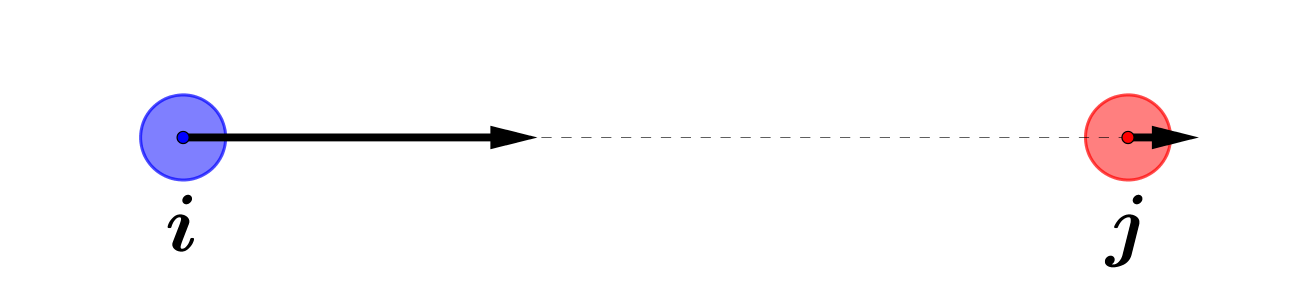}\label{fig:DefS1}}
    \subfigure[]{\includegraphics[width=0.5\linewidth]{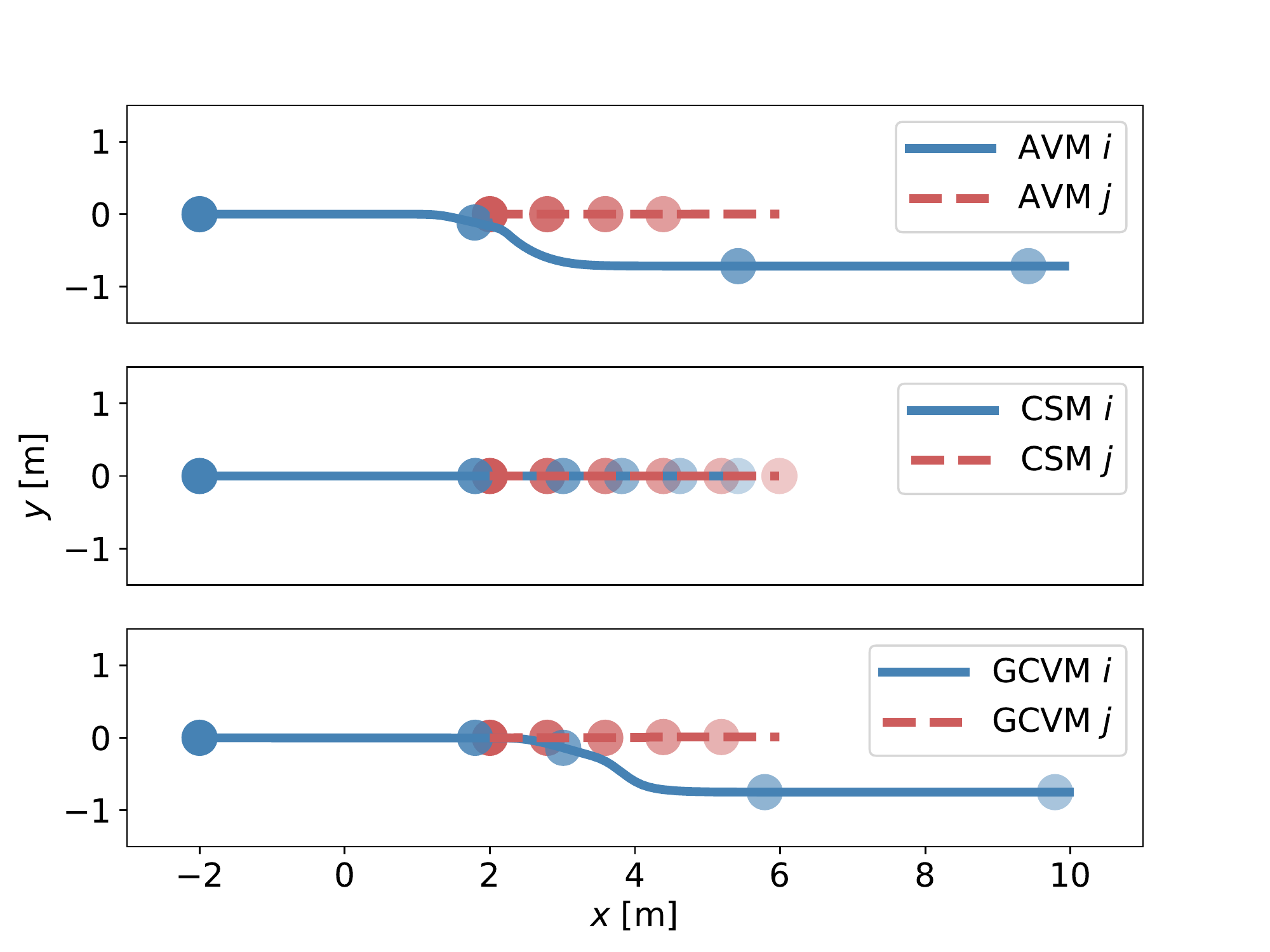}\label{fig:TrajS1}}
    \caption{
    (a): Scenario 1, agent $i$ walks behind agent $j$. 
    The two agents have the same desired direction but the free speed of agent $i$ is higher than that of agent $j$.
    (b): The trajectory of agents. The position of agents at different times is represented by the disk, and the transparency of these disks increases with increasing time. 
    When the AVM is used, overtaking starts earlier. 
    }
    \label{fig:scenario1}
\end{figure}

The first scenario is that agent $i$ walks behind agent $j$, which is shown in Figure~\ref{fig:DefS1}.
The agents have the same desired direction, but the free speed of agent $i$ is higher than that of agent $j$.
The trajectory of the agents in the first scenario is shown in Figure~\ref{fig:TrajS1}.
In the simulation using the GCVM and the AVM, agent $i$ overtakes agent $j$ by adjusting the direction of movement.
In the CSM, however, no overtaking is observed.
Moreover, compared to the GCVM, the overtaking in the AVM occurs earlier.

\begin{figure}[H]
    \centering
    \subfigure[]{\includegraphics[width=0.5\linewidth]{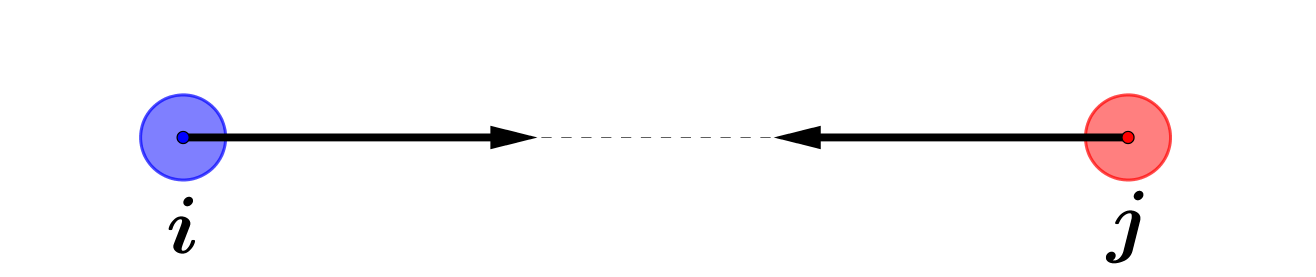}\label{fig:DefS2}}
    \subfigure[]{\includegraphics[width=0.5\linewidth]{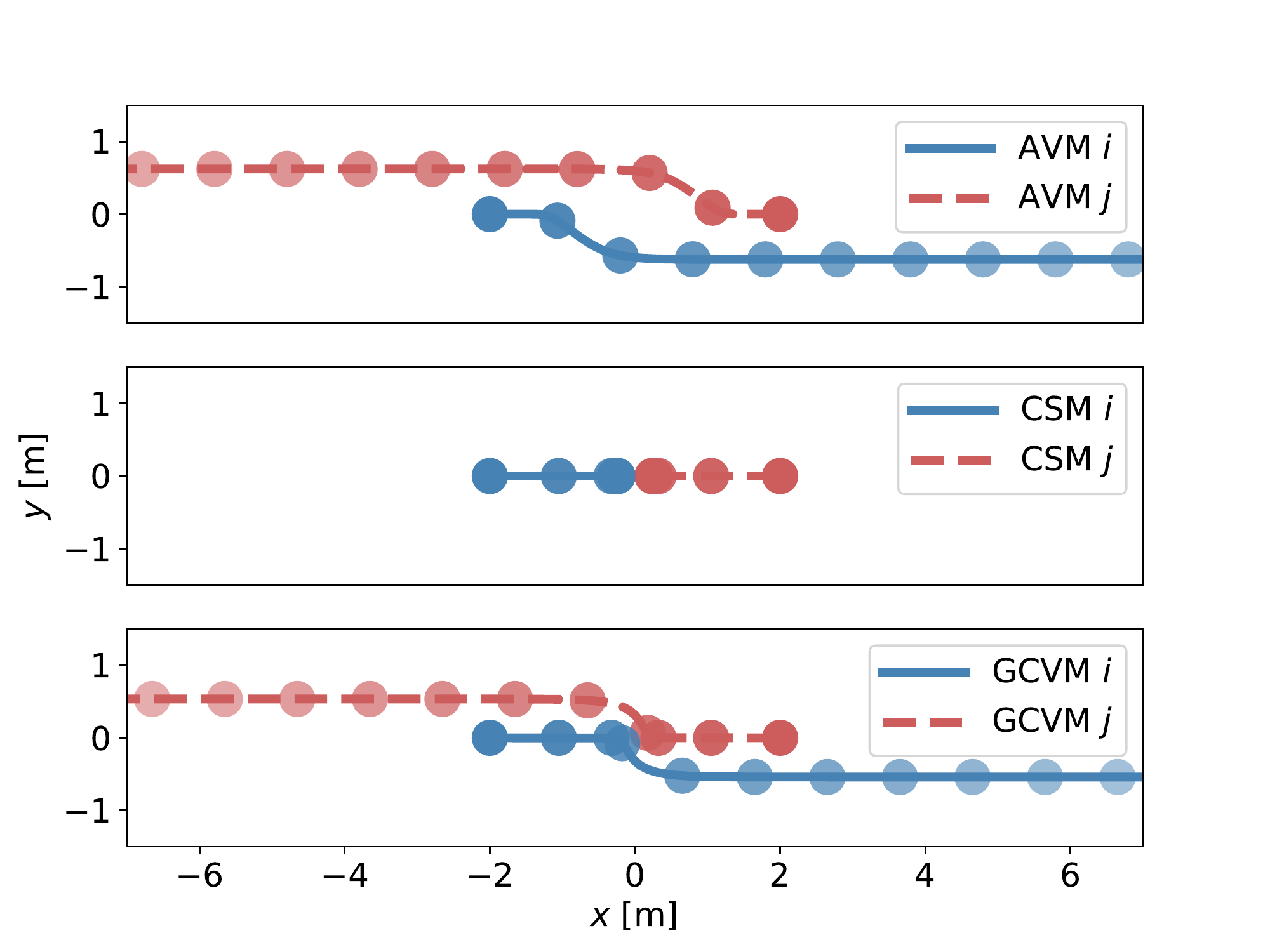}\label{fig:TrajS2}}
    \caption{
    (a): Scenario 2, agent $i$ and agent $j$ move toward each other.
    The two agents have the same free speed but opposite desired directions.
    (b): The trajectory of agents. The position of agents at different times is represented by the disk, and the transparency of these disks increases with increasing time. 
    Evasive movement starts earlier when the AVM is used.
    }
    \label{fig:scenario2}
\end{figure}

The second scenario, more relevant to bidirectional flow, depicts two agents having the same free speed with opposite desired directions (see Figure~\ref{fig:DefS2}).
The trajectory of the agents in the second scenario is shown in Figure \ref{fig:TrajS2}. 
Here again, it is observed that with the GCVM and the AVM, agents $i$ and $j$ both change their paths to avoid the imminent conflict, although this maneuver occurs earlier in the AVM than in the GCVM.
In the CSM, the two agents are unable to pass each other.

\begin{figure}[H]
    \centering
    \subfigure[]{\includegraphics[width=0.2\linewidth]{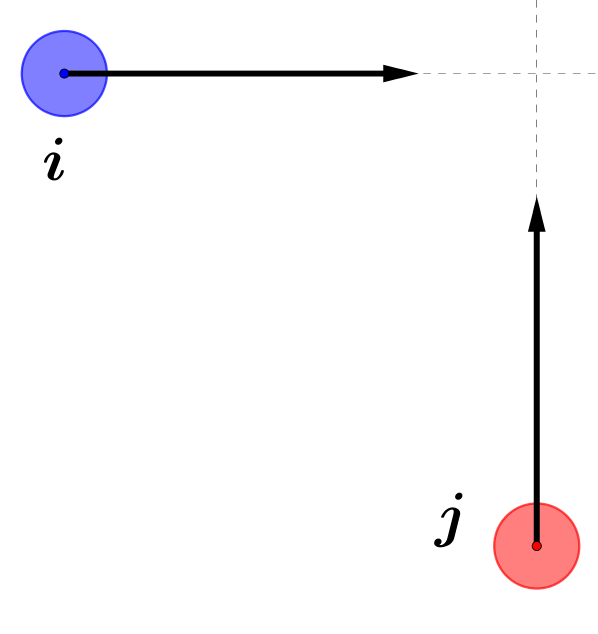}\label{fig:DefS3}}
    \subfigure[]{\includegraphics[width=0.75\linewidth]{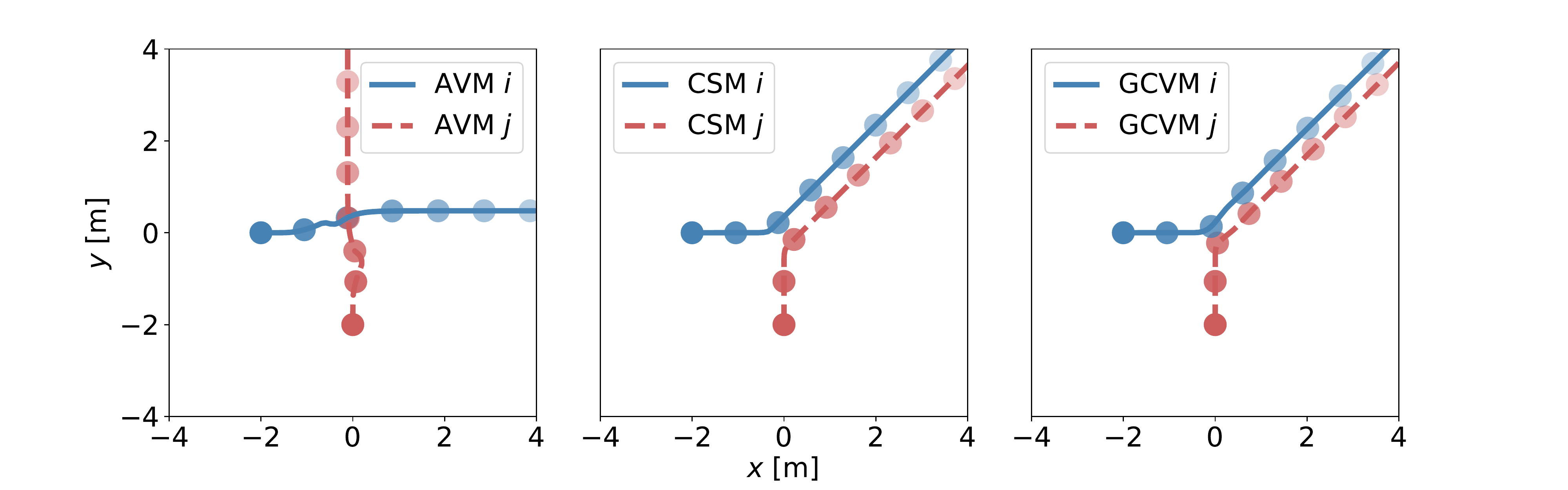}\label{fig:TrajS3}}
    \caption{
    (a): Scenario 3, agent $i$ and agent $j$ move across each other's path.
    The free speeds of the two agents are very close but not equal and their desired directions are perpendicular to each other.
    (b): The trajectory of agents. The position of agents at different times is represented by the disk, and the transparency of these disks increases with increasing time.
    }
    \label{fig:scenario3}
\end{figure}

In the last scenario (see Figure~\ref{fig:DefS3}), the paths cross at right angles.
The free speeds of the two agents are very similar but not quite equal to avoid the symmetric movement of the two agents.
The trajectory of the agents in the last scenario is shown in Figure~\ref{fig:TrajS3}.
When the AVM is used, the agents deviate slightly from the desired direction to the target and avoid collision.

In all cases, it can be concluded that, without introducing noise terms into the models, the movement of agents in the simulation using the AVM is closer to reality than using the CSM and the GCVM, where agents have difficulty overtaking or performing realistic evasive movements. 

\section{Bidirectional flow simulations with periodic boundary conditions}
\label{sec:bidirection}
\subsection{States of bidirectional flow}
\label{sec:states}
\begin{figure}[H]
    \centering
    \includegraphics[width=1.0\linewidth]{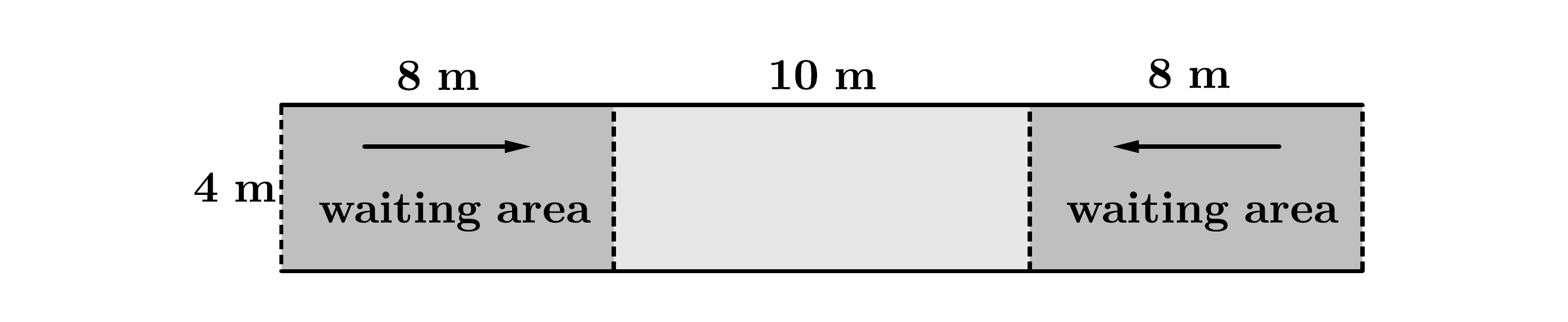}
    \caption{The corridor for bidirectional flow simulations.}
    \label{fig:geo}
\end{figure}

The bidirectional flow simulation is performed for a corridor shown in Figure~\ref{fig:geo}.
For the initial conditions of a simulation, the agents are randomly distributed in the gray waiting areas. 
The same number of agents were placed at the left and the right sides of the corridor.
After the simulation starts, agents in the left waiting area move toward the right, and vice versa.
Different initial conditions of agents' desired direction $\vec{e}^{~0}$ were compared before performing the simulations in this section.
Since no significant difference could be observed between the simulation results of ordered and unordered initial conditions, agents' desired direction $\vec{e}^{~0}$ were set parallel to the horizontal walls of the corridor.
The free speeds of agents are normally distributed $N\sim(1.55,0.18^2)~\SI{}{\meter\per\second}$ according to~\cite{zhang2012ordering}.
All simulations presented in this section are performed with periodic boundary conditions in the walking direction of the agents.
Each simulation lasts \SI{400}{\second}.

\begin{figure}[H]
    \centering
    \subfigure[]{\includegraphics[width=0.8\linewidth]{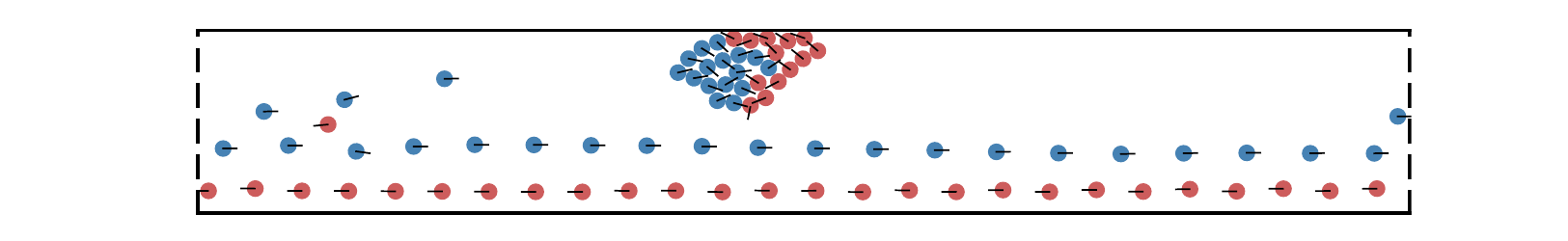}\label{fig:LocalJam}}
    \subfigure[]{\includegraphics[width=0.8\linewidth]{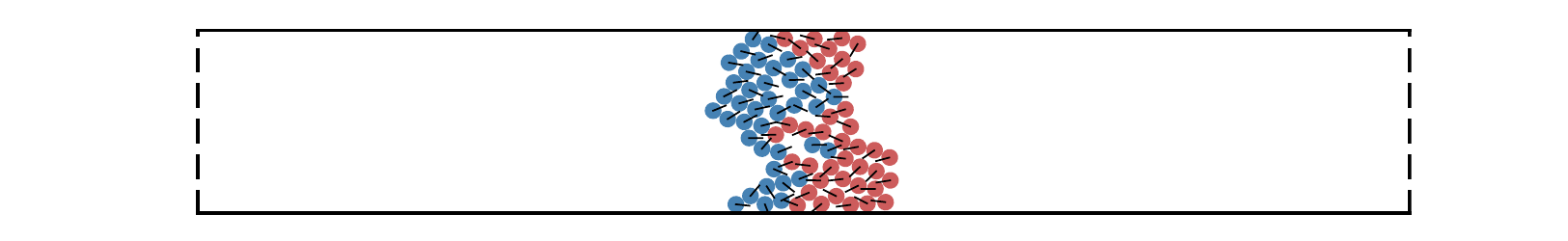}\label{fig:GlobalJam}}
    \subfigure[]{\includegraphics[width=0.8\linewidth]{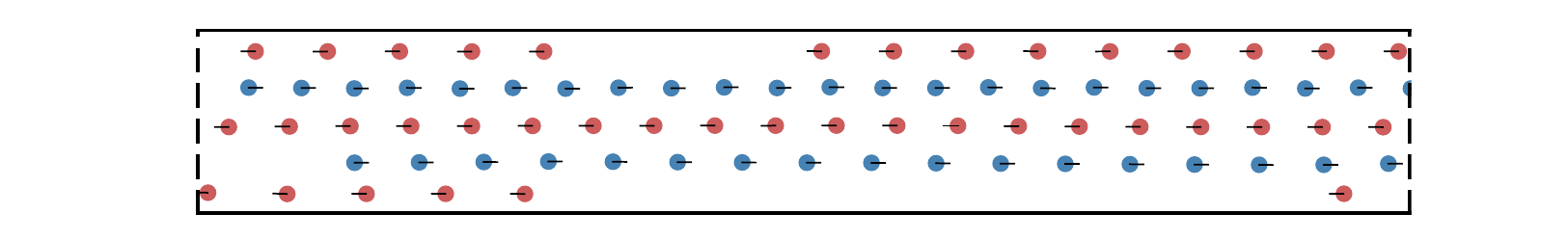}\label{fig:LaneFormation}}
    \subfigure[]{\includegraphics[width=0.8\linewidth]{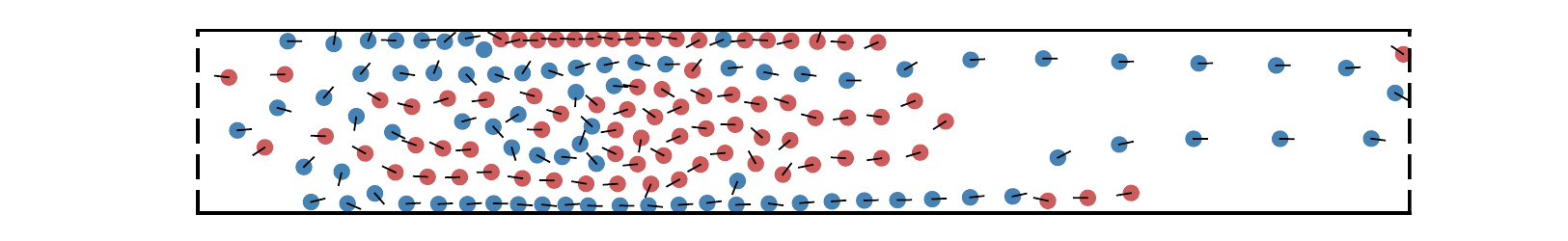}\label{fig:Disorder}}
    \caption{ Different states of bidirectional flow simulations with periodic boundary conditions.
    (a): Local jamming.
    (b): Global jamming.
    (c): Lane formation.
    (d): Disorder.}
    \label{fig:JamAndFluent}
\end{figure}

Based on the approach used in~\cite{nowak2012quantitative}, the patterns emerging in our simulations are classified into four different states, which are local jamming, global jamming, lane formation, and disorder (see Figure~\ref{fig:JamAndFluent}).
Local jamming and global jamming are both categorized as jamming states, whereas lane formation and disorder are grouped into the moving states.
If the average speed of an agent over \SI{10}{\second} is less than $v^0/100$, it is indicated as a static agent. 
Using this definition, a simulation is considered to be in the jamming state when the number of static agents in the simulation is equal to or greater than 2 ($N_\text{static}\ge2$); otherwise, the simulation is considered to be in a moving state.
Note, although \SI{400}{\second} is usually long enough to reach the steady state of the simulation, the jamming or moving state of a simulation still has a certain probability of being transient.
The number of static agents in the four simulations in Figure~\ref{fig:JamAndFluent} are 29 (local jamming), 80 (global jamming), 0 (lane formation), and 0 (disorder), respectively. 

In this section (Figures~\ref{fig:JamPDensity},~\ref{fig:JamkD}, and~\ref{fig:OPCompare}), each simulation is performed for $M=30$ times with different distributions of agents in the waiting areas, then the jamming probability $P_\text{jam}$ is calculated as
\begin{equation}
\label{equ:Pjam}
    P_\text{jam}=S_\text{jam}/M,
\end{equation}
where $S_\text{jam}$ is the number of simulations leading to a jamming state.

To further distinguish between the states of lane formation and disorder, the quantity $\Phi$ defined in \cite{von2019development} is introduced as 
\begin{equation}
\label{equ:GlobalOrder}
\Phi=\frac{1}{N} \sum_{i=1}^N \phi_i,
\end{equation}
with 
\begin{equation}
\label{equ:OrderAgent}
\phi_i=\frac{(N_i^\text{same}-N_i^\text{diff})^2}{(N_i^\text{same}+N_i^\text{diff})^2}\; \in [0,1],
\end{equation}
where $N_i^\text{same}$ is the set of all agents initially in the same waiting area as agent $i$ and currently moving in $i$'s lane and $N_i^\text{diff}$ is the set of all agents initially in a different waiting area to agent $i$ and currently moving in $i$'s lane.
The expressions of $N_i^\text{same}$ and $N_i^\text{diff}$ are 
\begin{equation}
\label{equ:SameLane}
N_i^\text{same}=\Big\{j, |y_j-y_i|<3r/2\ \text{and}\ \vec{e}_i^{~0} \cdot \vec{e}_j^{~0} >0 \Big\},
\end{equation}
\begin{equation}
\label{equ:DiffLane}
N_i^\text{diff}=\Big\{j, |y_j-y_i|<3r/2\ \text{and}\ \vec{e}_i^{~0} \cdot \vec{e}_j^{~0} <0 \Big\},
\end{equation}
where $y_i$ is the vertical position of agent $i$.

$\Phi$ is an indicator of how pronounced lanes are formed in a simulation. 
Knowing that an order parameter is only measured at steady states, $\Phi$ is used here to compare models with respect to their performance of describing the transient state of lane formation. 
The values of $\Phi$ in the four simulations in Figure~\ref{fig:JamAndFluent} are 0.66 (local jamming), 0.18 (global jamming), 1.00 (lane formation), and 0.31 (disorder), respectively.  
Nevertheless, there is no specific boundary to clearly distinguish between the states of lane formation and disorder.

\subsection{Jamming transition}
\label{sec:samePara}
To study the jamming transition in bidirectional flow, simulations are performed with the AVM, the CSM, and the GCVM, respectively.
The parameters of models are shown in Table~\ref{tab:parameterCompare}.
For each model, simulations are performed with different numbers of agents ranging from 20 to 200 (10 to 100 in each waiting area).

\begin{figure}[H]
    \centering
    \includegraphics[width=0.5\linewidth]{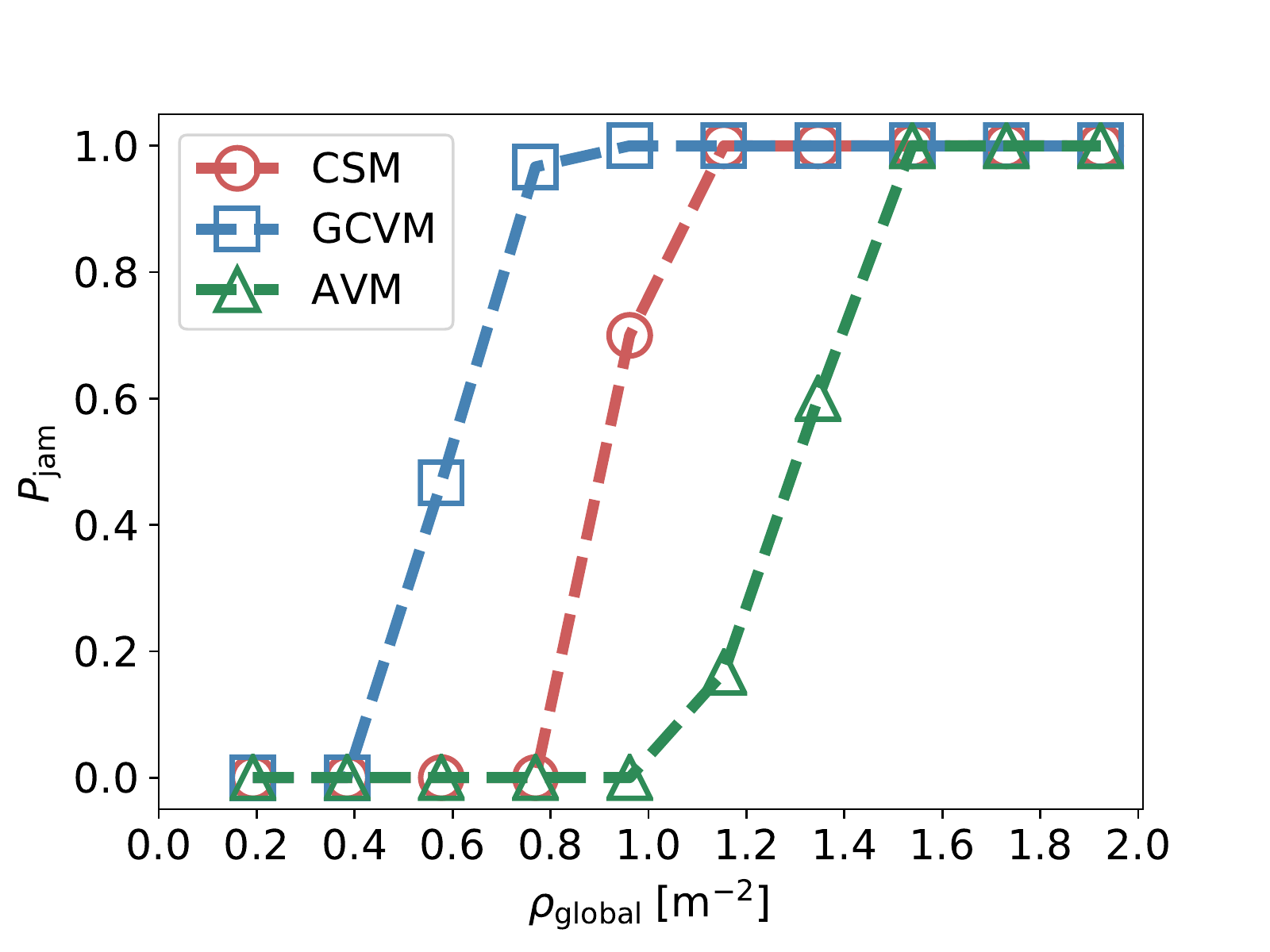}
    \caption{
    The relationship between $P_\text{jam}$ and $\rho_{\text{global}}$ for different models.
    }
    \label{fig:JamPDensity}
\end{figure}

The relationship between $P_\text{jam}$ and $\rho_{\text{global}}$, the global density of agents in the corridor, for different models is shown in Figure~\ref{fig:JamPDensity}.
The value of $\rho_{\text{global}}$ is equal to the number of agents divided by the area of the corridor.
With an increase in $\rho_{\text{global}}$, a transition from moving states ($P_\text{jam}=0$) to jamming states ($P_\text{jam}=1$) is observed in the simulation of all three models.  
However, with the AVM, the transition occurs at a higher value of $\rho_{\text{global}}$ compared to the other two models, which indicates the model's ability to reproduce lane formation even at higher density values.

\subsection{Parametric study}
\label{sec:diffPara}
Parameters $k$ and $D$ are used to calibrate the strength and range of the impact from neighbors to the direction of movement in all three models (the CSM, the GCVM, and the AVM).
Although the definitions vary slightly between the three models, higher values of $k$ and larger $D$ always led to agents being more stimulated to deviate from their desired directions.
In this section, the effect of $k$ and $D$ on the jamming probability $P_{\text{jam}}$ is studied for each of the three models.

For each model, simulations are performed with different values of k (1, 2, 3, 4, 5, and 6) and different values of D (0.01, 0.02, 0.05, 0.10, \SI{0.20}{\meter}).
To ensure simulations with jamming and moving states, the global density of agents $\rho_{\text{global}}$ is set close to the critical density between the moving state and the jamming state (see Figure~\ref{fig:JamPDensity}).
The number of agents is 100 ($\rho_{\text{global}}\approx\SI{0.96}{\per\square\meter}$) for the simulations using the CSM, 60 ($\rho_{\text{global}}\approx\SI{0.58}{\per\square\meter}$) for the simulations using the GCVM, and 140 ($\rho_{\text{global}}\approx\SI{1.35}{\per\square\meter}$) for the simulations using the AVM.
Other parameters of the three models are given in  Table~\ref{tab:parameterCompare}.

\begin{figure}[H]
    \centering
    \subfigure[]{\includegraphics[width=0.3\linewidth]{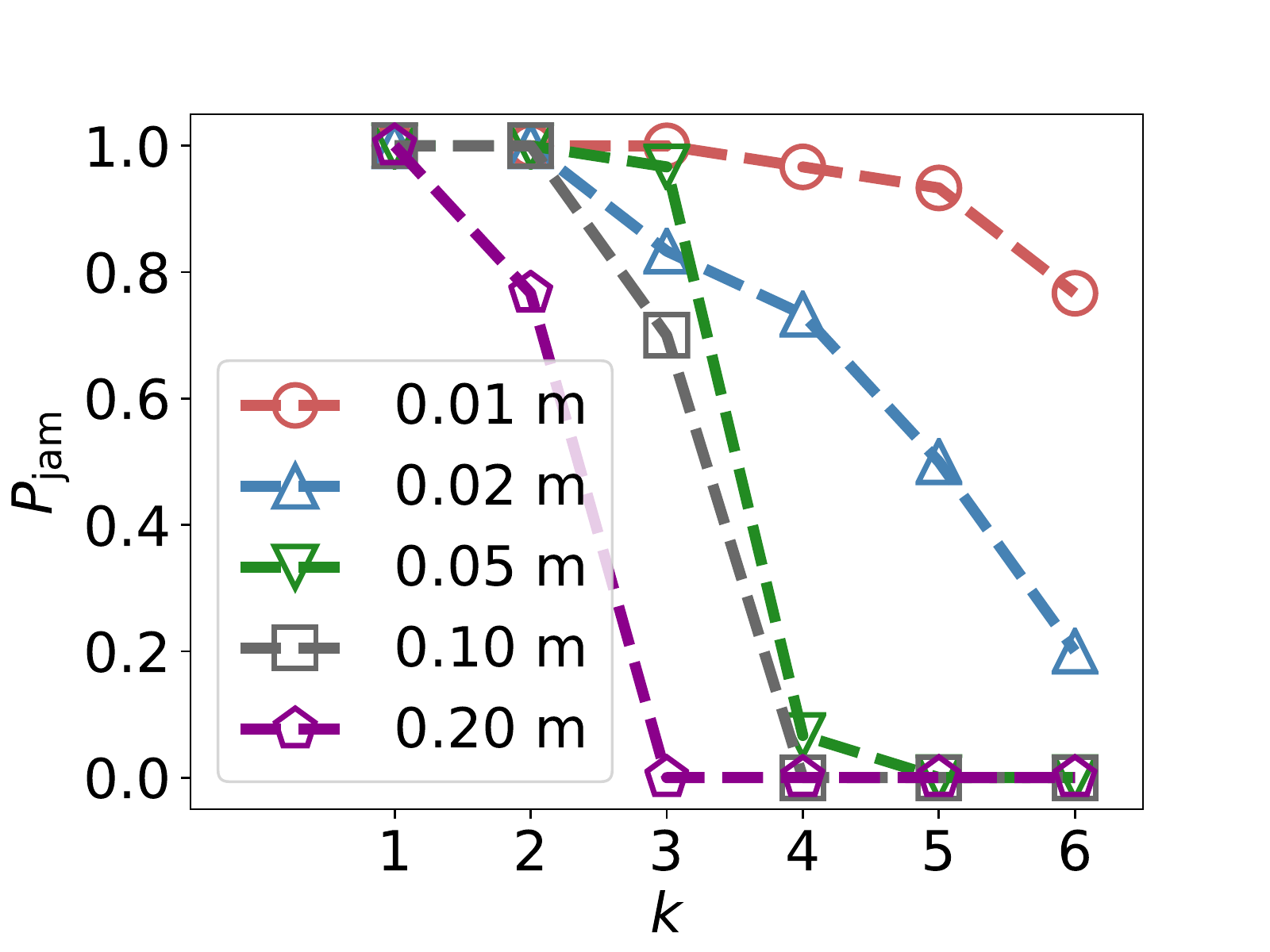}\label{fig:JamPkDCSM}}
    \subfigure[]{\includegraphics[width=0.3\linewidth]{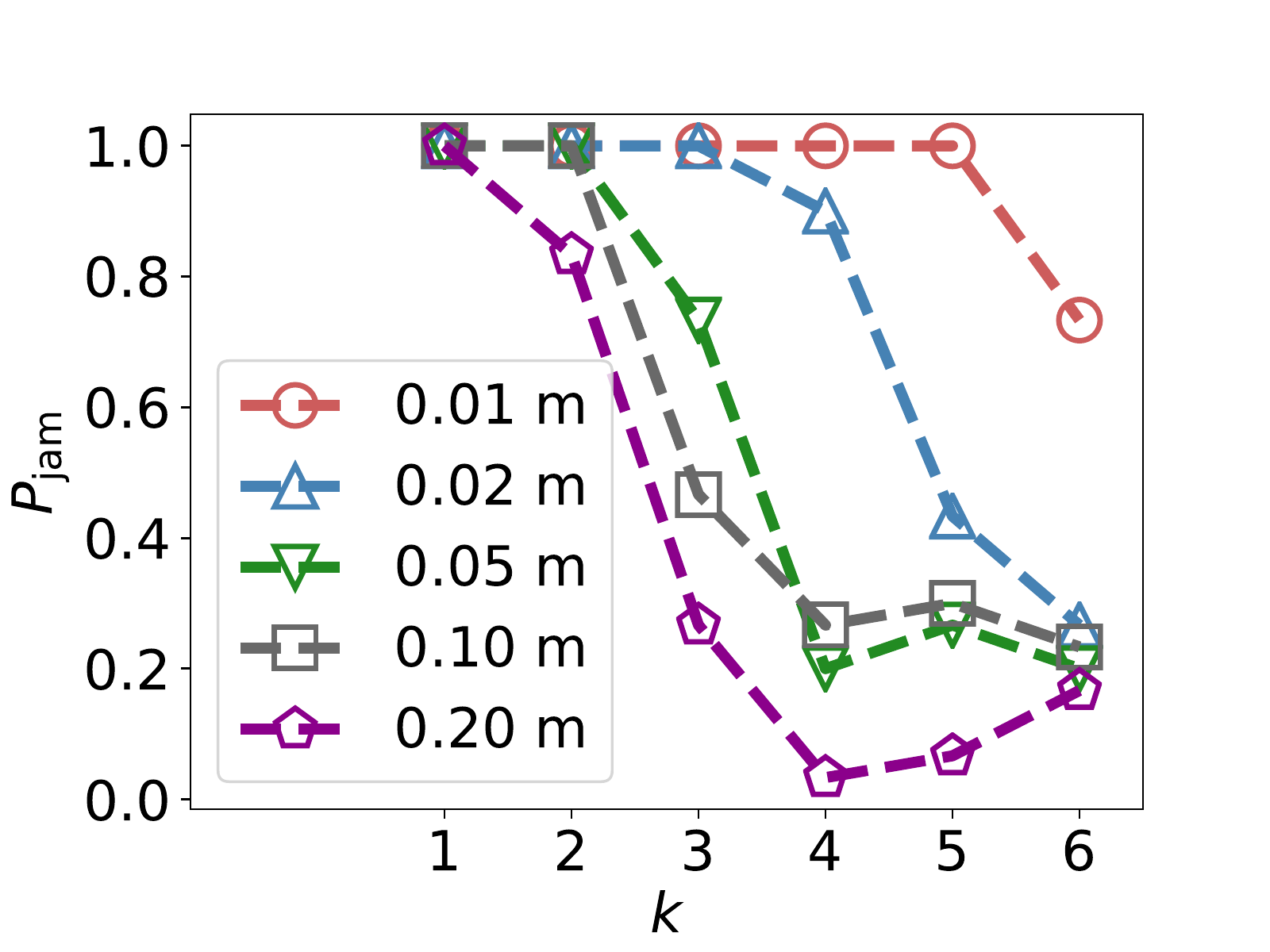}\label{fig:JamPkDGCVM}}
    \subfigure[]{\includegraphics[width=0.3\linewidth]{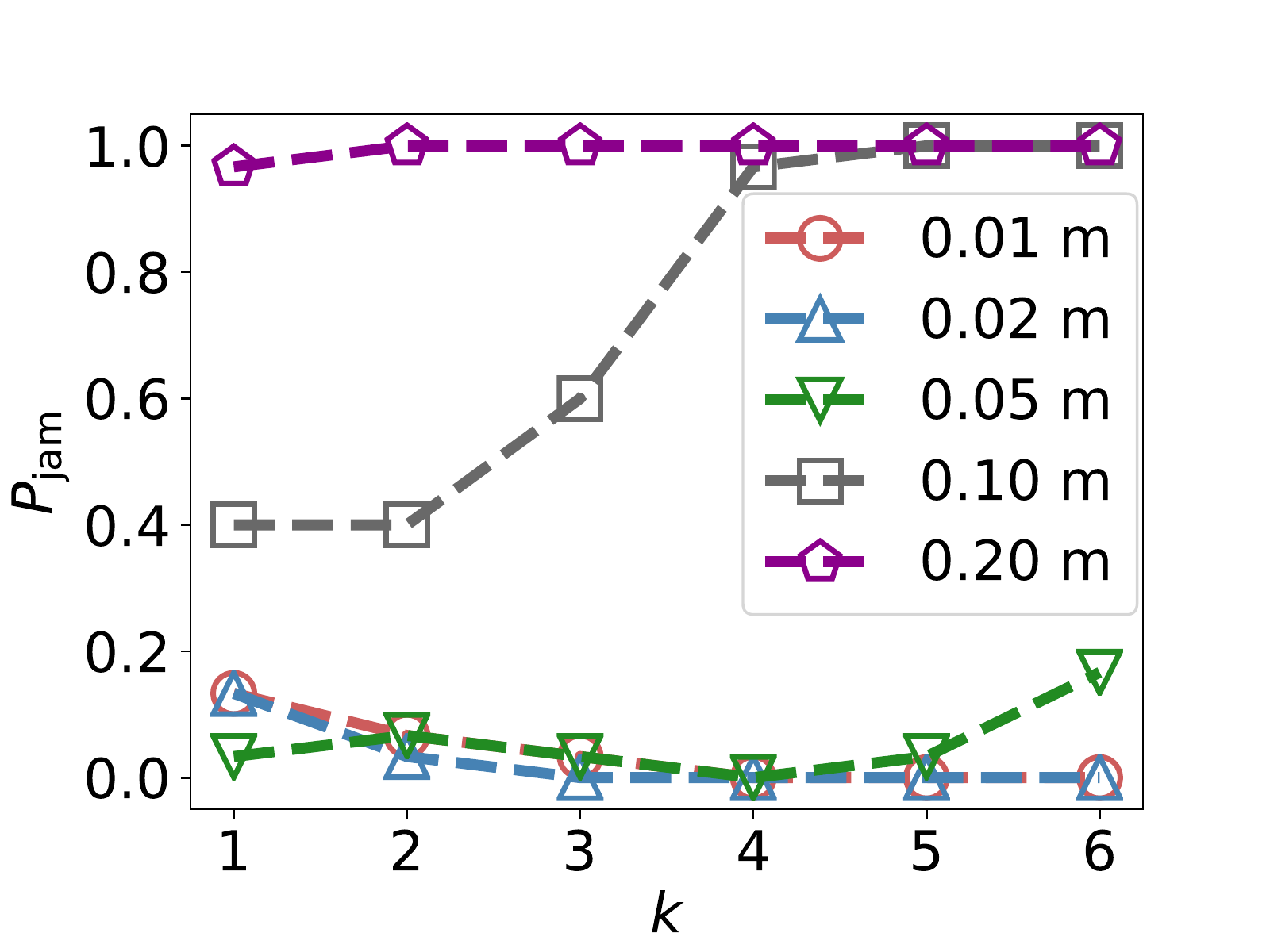}\label{fig:JamPkDAVM}}
    \caption{
    The relationship between $P_\text{jam}$ and $k$ for different values of $D$.
    The mean value of $D$ can be read in the legend.
    (a): CSM.
    (b): GCVM.
    (c): AVM.
    }
    \label{fig:JamkD}
\end{figure}

The jamming probability of the simulation using the CSM and the GCVM is shown in Figures~\ref{fig:JamPkDCSM} and~\ref{fig:JamPkDGCVM}, respectively.
Generally speaking, the value of $P_\text{jam}$ in the simulation using the CSM and the GCVM decreases with increasing $k$ and $D$, which means that the jamming probability decreases with increasing impact from neighbors on the direction of movement.
However, the value of $P_\text{jam}$ in the simulation using the AVM shows a different trend (see Figure~\ref{fig:JamPkDAVM}).
When the value of $D$ is small (0.01, 0.02, and \SI{0.05}{\meter}), the value of $P_\text{jam}$ changes slightly.
With a larger value of $D$ (\SI{0.1}{\meter}), the value of $P_\text{jam}$ increases with increasing $k$.   
Note that when the value of $D$ is large enough (\SI{0.2}{\meter}), the value of $P_\text{jam}$ is close to 1 and the effect of $k$ is marginal again.
Generally, in the AVM, with increasing impact from neighbors on the direction of movement, the jamming probability increases.

On the basis of all combinations of $k$ and $D$ in Figure~\ref{fig:JamkD}, the set of parameters that leads to the minimal $P_\text{jam}$ is identified for each model and shown in Table~\ref{tab:OPParameter}.

\begin{table}[H]
    \centering
    \begin{tabular}{|l|c|c|c|}
    \hline
             & CSM & GCVM & AVM   \\
    \hline
         $k$ & 6 & 4 & 6\\
    \hline
         $D$ [\SI{}{\meter}] & 0.2 & 0.2 & 0.01\\
    \hline
    \end{tabular}
    \caption{The set of $k$ and $D$ that leads to the minimal $P_\text{jam}$ in Figure~\ref{fig:JamkD}.}
    \label{tab:OPParameter}
\end{table}

\subsection{Lane formation}
\begin{figure}[H]
    \centering
    \subfigure[]{\includegraphics[width=0.45\linewidth]{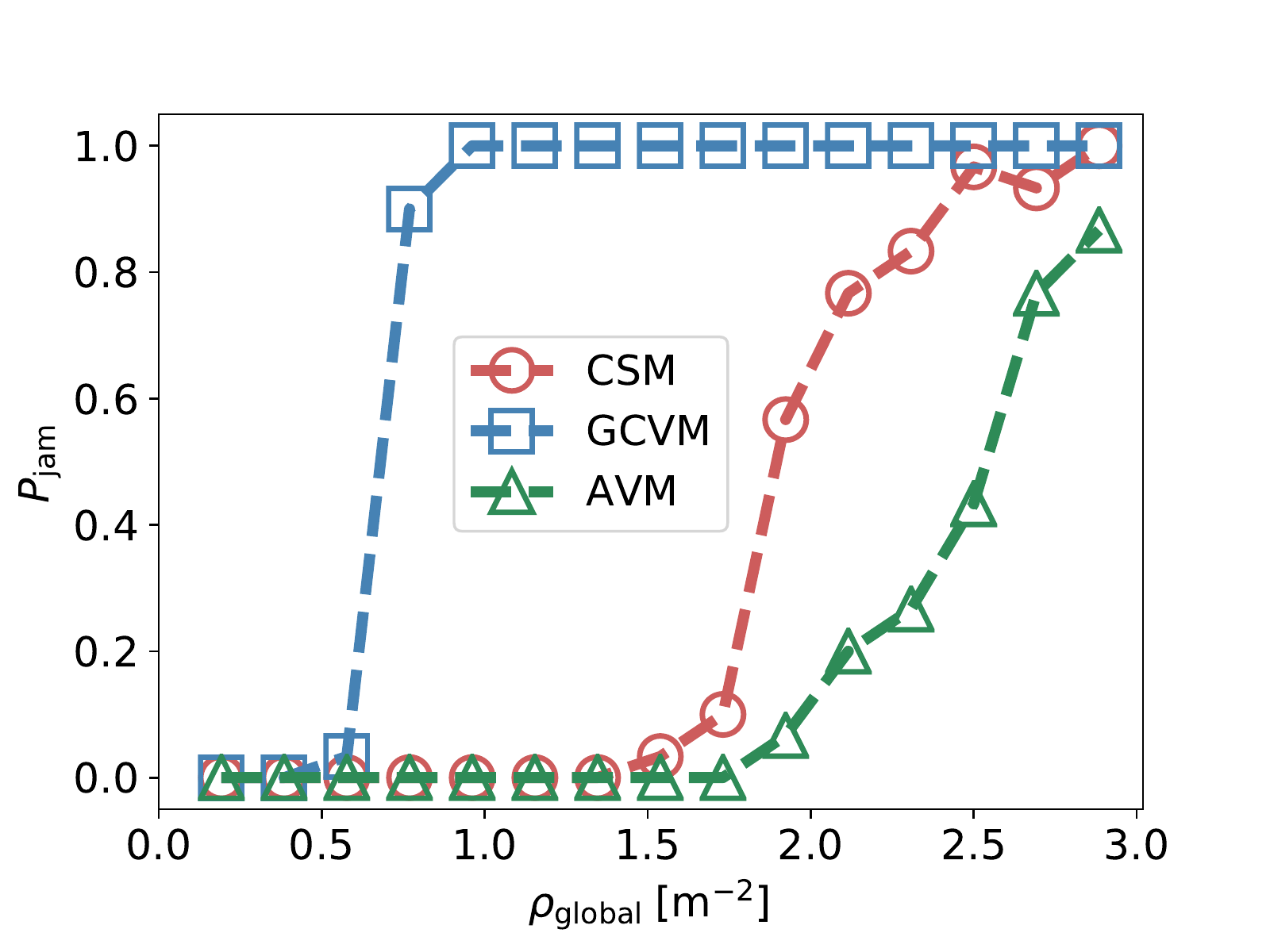}\label{fig:OPCompare}}
    \subfigure[]{\includegraphics[width=0.45\linewidth]{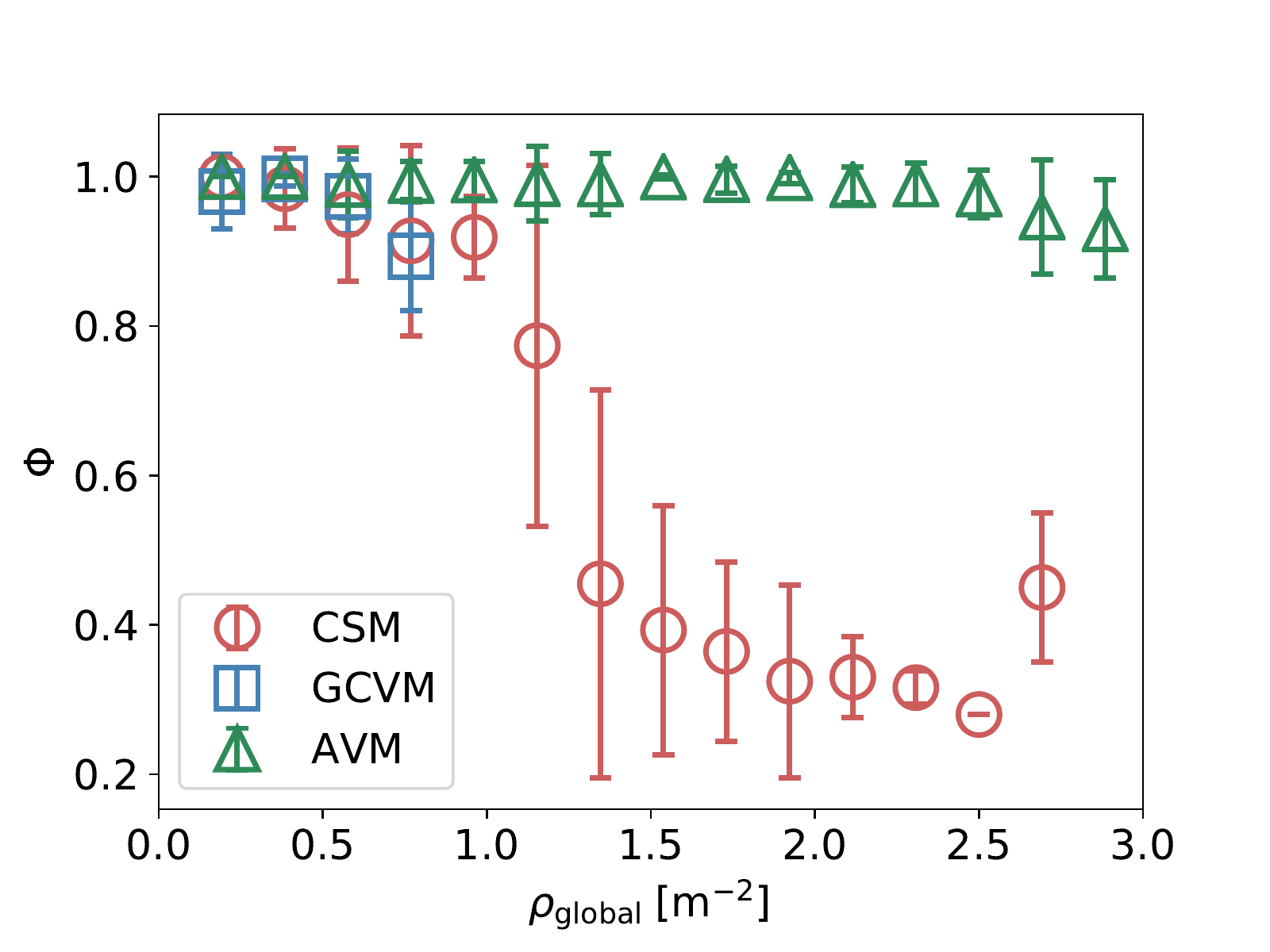}\label{fig:OrderDensityOP}}
    \caption{
    Using the values of $k$ and $D$ from Table~\ref{tab:OPParameter}.
    (a): The relationship between $P_\text{jam}$ and $\rho$ for different models.
    (b): The mean value and standard deviation of $\Phi$ in the simulations with the moving state, for each model and each density.
    }
    \label{fig:OP}
\end{figure}

The values of $k$ and $D$ from Table~\ref{tab:OPParameter}, together with other parameter values from Table~\ref{tab:parameterCompare}, are used for a comparative study of lane formation in the three models.
For each model, simulations are performed with different numbers of agents from 20 to 300 ($\rho_{\text{global}}$ from 0.19 to \SI{2.8}{\per\square\meter}).
The relationship between $P_\text{jam}$ and $\rho_{\text{global}}$ for different models is shown in Figure~\ref{fig:OPCompare}.
Compared to Figure~\ref{fig:JamPDensity}, the jamming probability $P_\text{jam}$ is significantly reduced in the simulation using the CSM and the AVM by adopting the optimal values of $k$ and $D$. 
The GCVM, however, does not show any improvement.

Furthermore, to gain a better insights into the lane formation phenomenon, the average value of $\Phi$ in the last \SI{10}{\second} of each simulation is calculated.
First, simulations are classified as ``jamming'' or ``moving'' according to their states.
Then, the mean value and standard deviation of $\Phi$ in the simulations with the moving state are calculated for each model and each density.
The variations in $\Phi$ with respect to the global densities are shown in Figure~\ref{fig:OrderDensityOP}.
For all the three models, when $\rho_{\text{global}}<\SI{1.0}{\per\square\meter}$, the values of $\Phi$ in the moving states are always close to 1, which indicates that, in this case, lanes are always formed. 
When $\rho_{\text{global}}>\SI{1.0}{\per\square\meter}$, for the moving states, the value of $\Phi$ is unavailable in the GCVM, and the values in the AVM are significantly higher than in the CSM.
This indicates that, although both are in the moving states, the AVM reproduces lane formation much better.

\begin{figure}[H]
    \centering
    \subfigure[]{\includegraphics[width=0.45\linewidth]{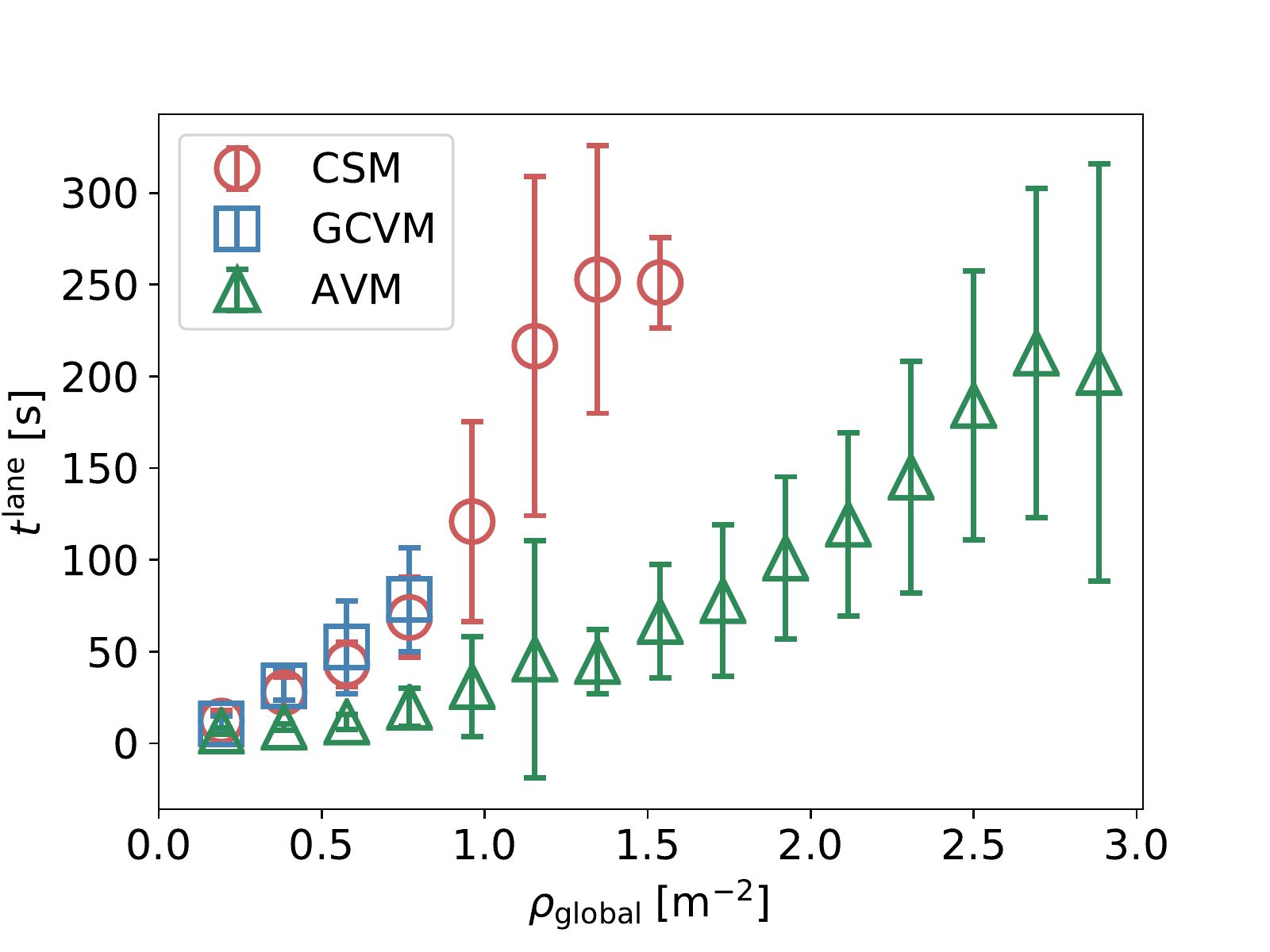}\label{fig:LaneTimeDensityOP}}
    \subfigure[]{\includegraphics[width=0.45\linewidth]{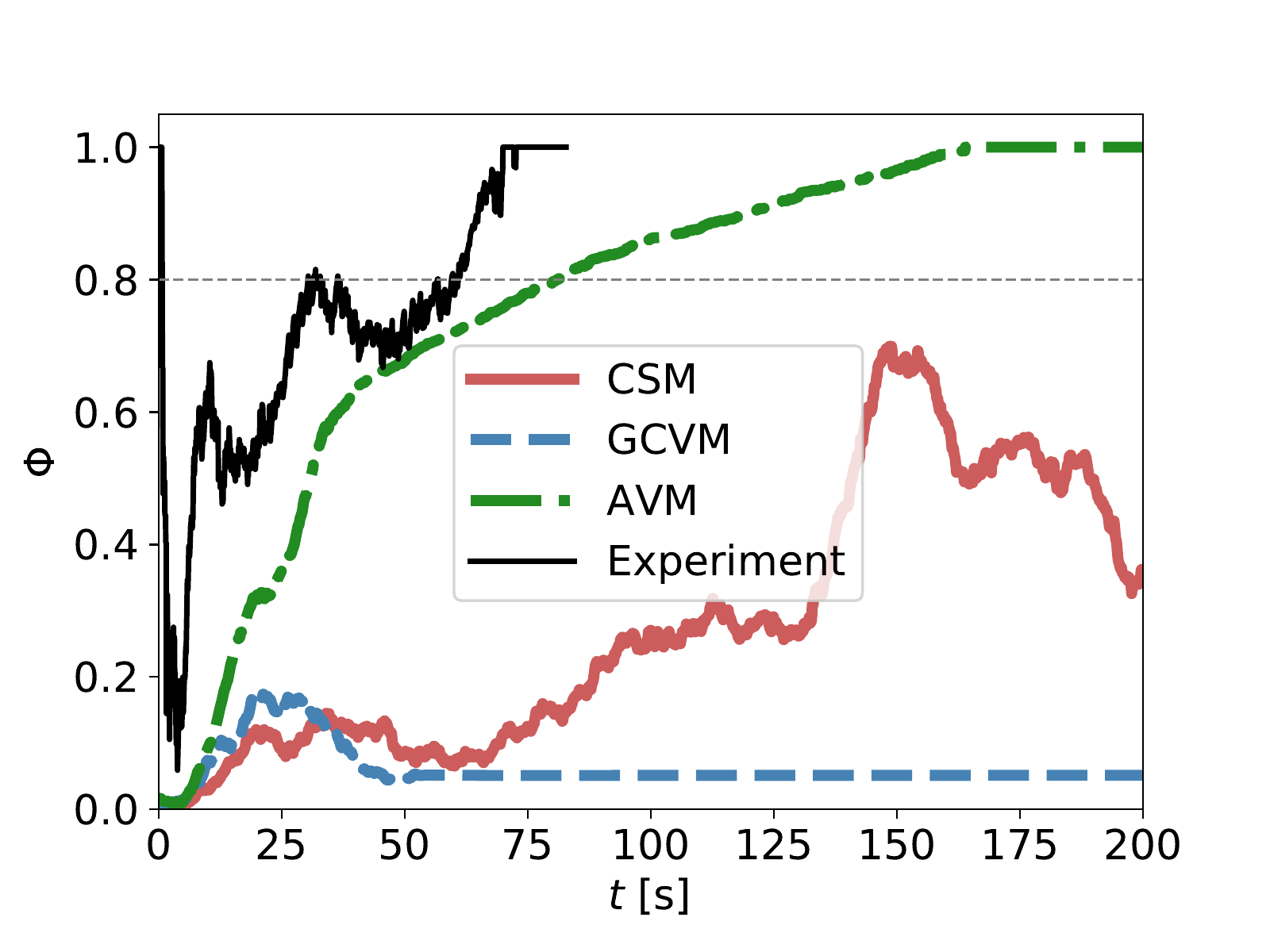}\label{fig:SingleOrderOP}}
    \caption{
    (a) The mean value and standard deviation of $t^{\text{lane}}$ in the simulations with the moving state, for each model and each density.
    (b):The relationship between the value of $\Phi$ and the simulation time $t$ of three simulations using different models ($\rho_{\text{global}}$=\SI{1.92}{\per\square\meter}) and an experiment~\cite{zhang2012ordering}.}
\end{figure}

The quantity $\Phi$ is a reliable indicator of the formation of lanes in a simulation. 
Since it changes in time, an artificial threshold of 0.8 for $\Phi$ is introduced to describe how fast lanes are developed.
The time when the value of $\Phi$ first exceeds the threshold is denoted by $t^{\text{lane}}$.
The mean value and standard deviation of $t^{\text{lane}}$ in simulations with the moving state are calculated for each model and each density.
See Figure~\ref{fig:LaneTimeDensityOP}.
For all three models, the mean value of $t^{\text{lane}}$ increases with increasing $\rho_{\text{global}}$, which is to be expected.
Moreover, lanes form much faster in the simulation using the AVM than with the other models.

\begin{figure}[H]
    \centering
    \includegraphics[width=\linewidth]{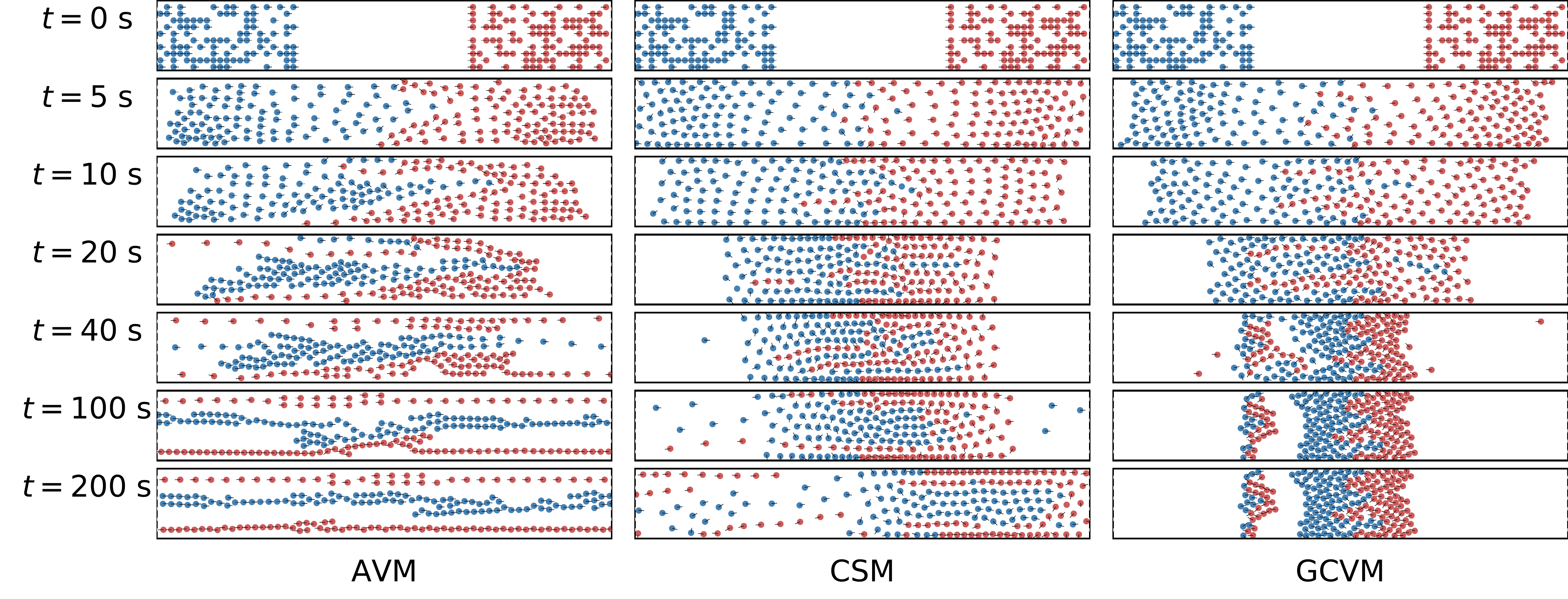}
    \caption{
    The snapshot of the simulations using the AVM, the CSM, and the GCVM (from left to right). 
    From top to bottom: $t=$ 0, 5, 10, 20, 40, 100, and \SI{200}{\second}.}
    \label{fig:framesModelsOP}
\end{figure}

Three single simulations using different models are further compared with bidirectional flow experiment~\cite{zhang2012ordering}.
Figure~\ref{fig:SingleOrderOP} shows the time series of $\Phi$ in the first \SI{200}{\second}.
Here, $\Phi$ continues to increase to 1 and then remains stable in the simulation using the AVM, while it keeps fluctuating below 0.7 for the CSM. 
For the GCVM, the quantity $\Phi$ is stable at a low value, indicating a lack of any lane formation.
The possible reason for the difference between the AVM and the CSM here is that the moving state of the AVM can be attributed to the strategy of following, while the moving state of the CSM is due to agents pushing each other aside.
This behavior is also reflected in the snapshots of the three simulations at different times, shown in Figure~\ref{fig:framesModelsOP}.
The experiment compared in Figure~\ref{fig:SingleOrderOP} is performed in a \SI{3.6}{\meter} corridor under open boundary conditions and it records the trajectories of pedestrians in the \SI{8}{\meter} length measurement area.
The steady density of pedestrians in the measurement area is around \SI{2}{\per\square\meter}.
The value $\Phi$ for pedestrians in the measurement area is calculated and the lane formation occurs even earlier in the experiment than in the simulation.
It could not be excluded that, in the experiment, the formation of lanes starts even outside of the measurement area. 
Thus, the comparability of experiments and simulations is limited.
However, a comparison of the time series of $\Phi$, in particular, the increase in $\Phi$ over time, provides a rough estimate whether the time in which the system switches from an unordered state without lanes to an ordered state with lanes has the same order of magnitude. 

\section{Validation of the AVM using the fundamental diagram}
\label{sec:experiment}
After the AVM was compared to two other models and showed its ability to produce lane formation reasonably, the model was then validated with respect to the fundamental diagram.
For this purpose, the fundamental diagram (FD) obtained from bidirectional flow experiments in~\cite{zhang2012ordering,cao2017fundamental} was used to calibrate the parameters of the AVM.
Bidirectional flow simulations with open boundary conditions were performed in a corridor shown in Figure~\ref{fig:geo}. 
The same number of agents were placed at the left and the right side of the corridor.
The number of agents is varied in different simulations to realize different local densities in the corridor. 
Agents in the left waiting area move toward the right, and vice versa.
The simulation ends when all agents leave the corridor.
The calibration was performed manually and resulted in the set of parameters listed in Table~\ref{tab:validatedPara}.

\begin{table}[H]
    \centering
    \begin{tabular}{|c|c|c|c|c|c|c|c|c|}
    \hline
    Flow types & $v^0~\rm[\SI{}{\meter\per\second}]$& $r$ [\SI{}{\meter}]& $k$&  $D$ [\SI{}{\meter}]& $T$ [\SI{}{\second}]&  $\tau$ [\SI{}{\second}]&  $t^{\text{a}}$ [\SI{}{\second}]\\
    \hline
    Bidirectional & \multirow{2}{*}{$N\sim(1.55,0.18^2)$}& \multirow{2}{*}{0.18}& \multirow{2}{*}{6}& \multirow{2}{*}{0.01}& 0.6& \multirow{2}{*}{0.3}& \multirow{2}{*}{0.75}\\
    \cline{1-1}\cline{6-6}
    Unidirectional & & & & & 0.5 & & \\
    \hline
    \end{tabular}
       \caption{
    Validated parameters of the AVM.
    }
    \label{tab:validatedPara}
\end{table}

\begin{figure}[H]
    \centering
    \subfigure[]{\includegraphics[width=0.4\linewidth]{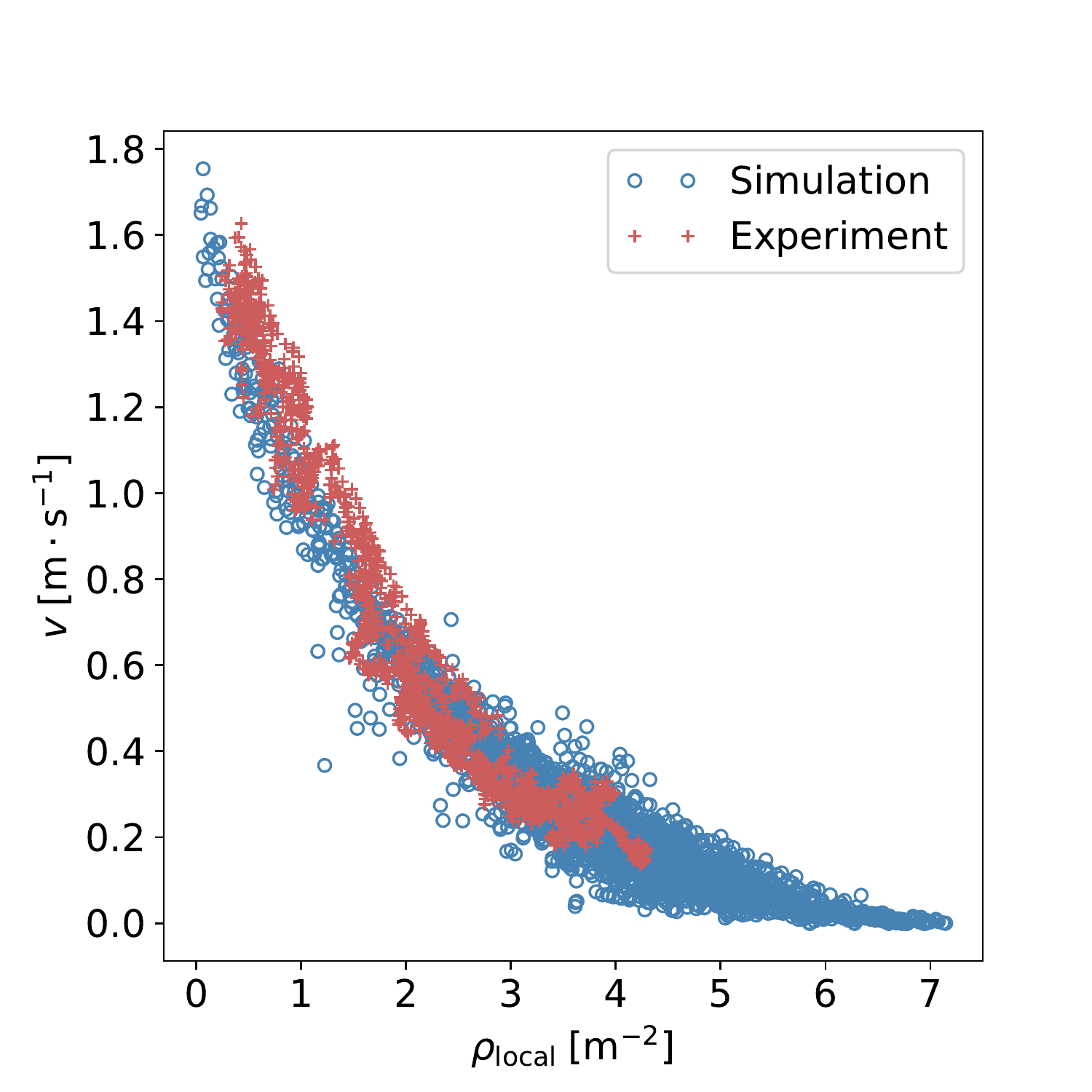}\label{fig:fdv}}
    \subfigure[]{\includegraphics[width=0.4\linewidth]{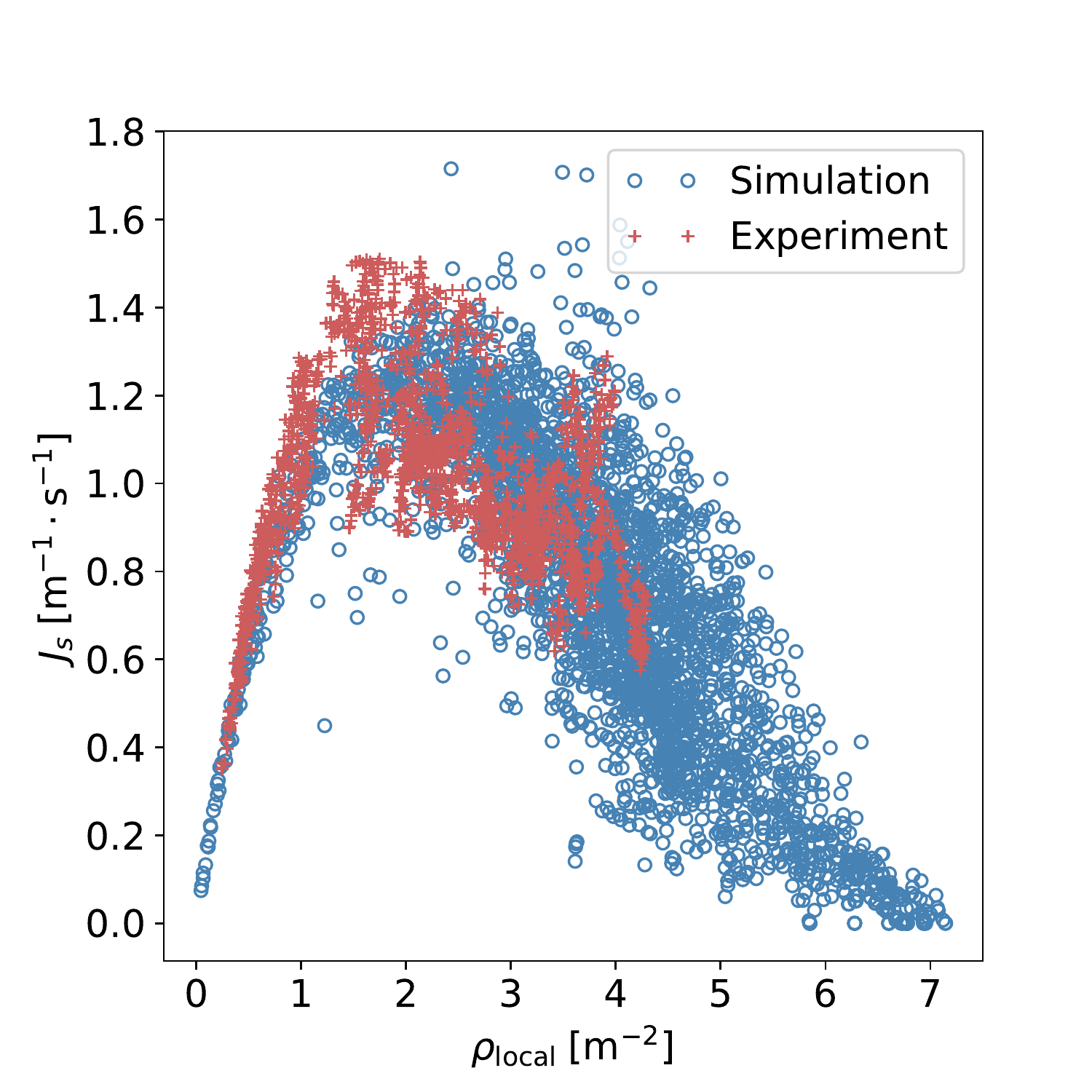}\label{fig:fdf}}
    \caption{The fundamental diagram of bidirectional flow from the experiment and the simulation.
    (a) Density-velocity.
    (b) Density-specific flow.}
    \label{fig:fd}
\end{figure}

The FD obtained by simulations and experiments are compared in~Figure~\ref{fig:fd}.
The speed $v$, the local density $\rho_\text{local}$, and the specific flow $J_s=\rho_\text{local} \cdot v$ in both the simulation and the experiment are measured from trajectories of pedestrians or agents using the same method.
The measurement method is proposed in~\cite{cao2017fundamental}, where the local density is measured using the Voronoi method and the measured speed is the projection of the real speed in the horizontal direction.
The FD obtained from the simulation is consistent with the FD obtained from the experiment.

\begin{figure}[H]
    \centering
    \includegraphics[width=0.6\linewidth]{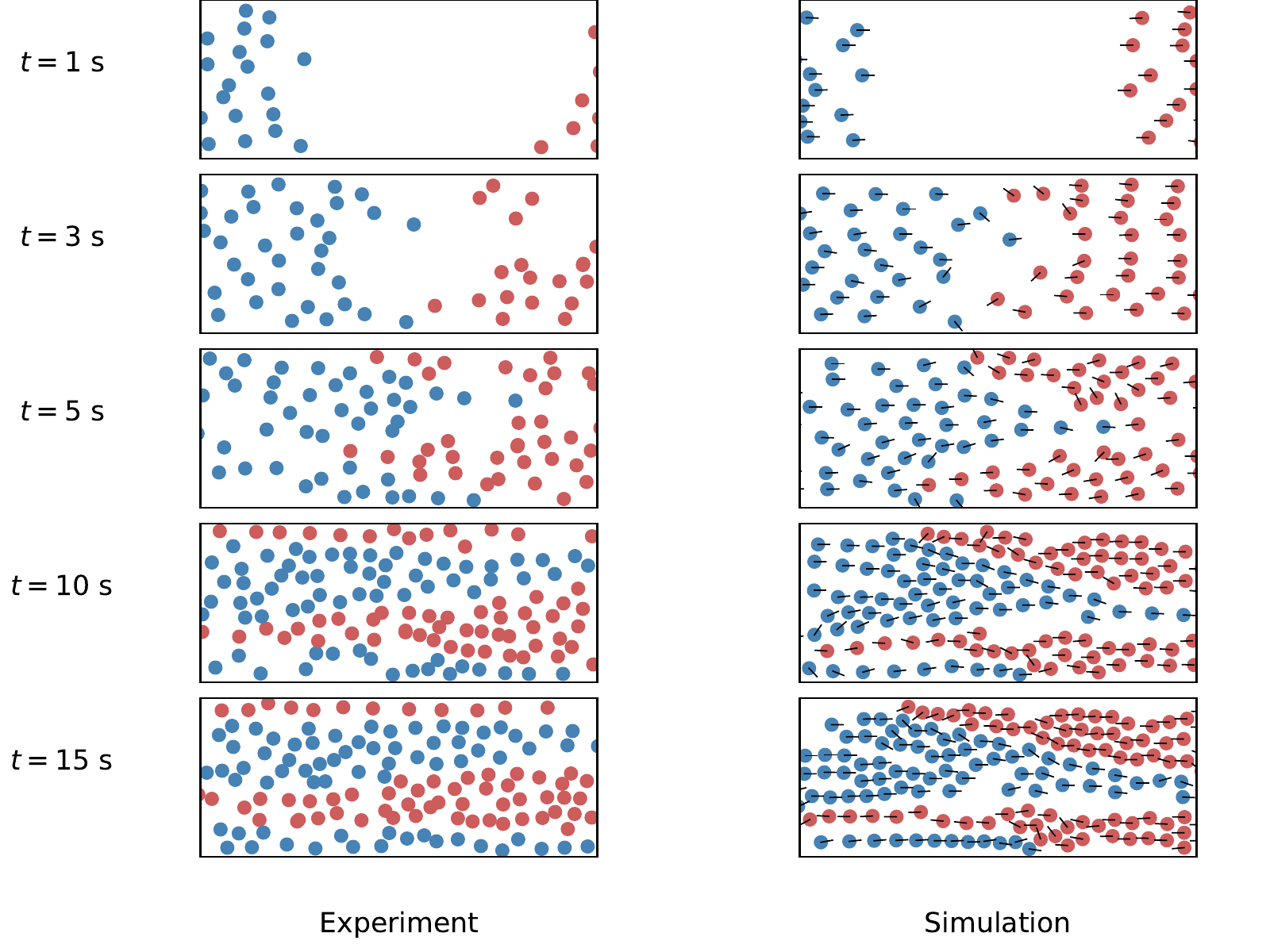}
    \caption{
    The trajectory snapshots of an experiment (left) and a simulation (right). 
    Top to bottom: $t=$ 1, 3, 5, 10, and \SI{15}{\second}.}
    \label{fig:framesExp}
\end{figure}

As well as the quantitative comparison of the FD, the process of lane formation in the simulation is qualitatively compared to the experiment (see Figure~\ref{fig:framesExp}).
The trajectory snapshots of a simulation and a bidirectional flow experiment in a \SI{4}{\meter} wide corridor are compared. 
Note that the trajectories of the experiment are superimposed by the swaying of the head movement due to the bipedal movement in steps, which is not covered by the model. 
Moreover, the course of lane formation in the simulation is similar in time to the experiment.

\begin{figure}[H]
    \centering
    \subfigure[]{\includegraphics[width=0.4\linewidth]{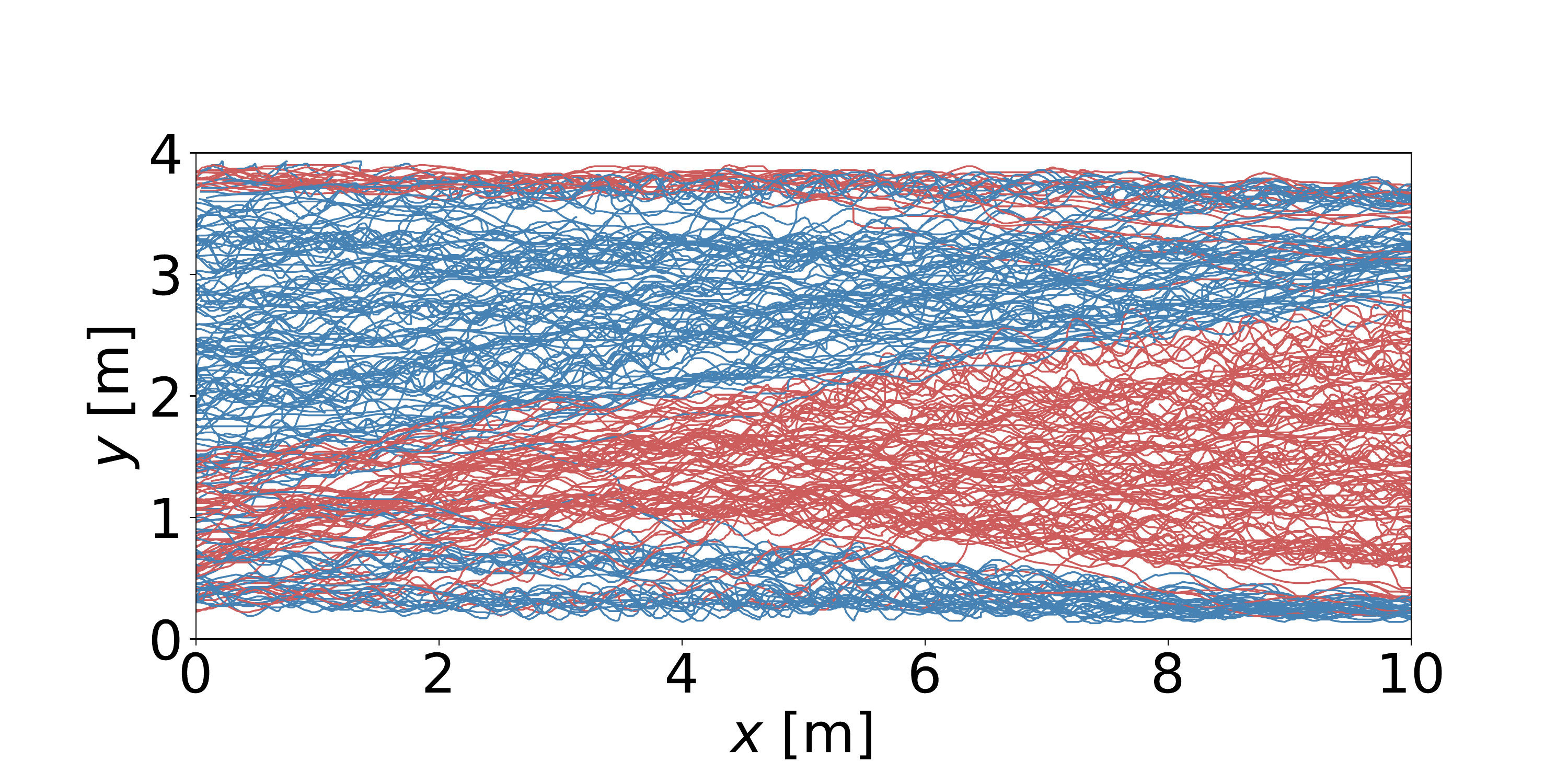}\label{fig:ExperimentTraj}}
    \subfigure[]{\includegraphics[width=0.4\linewidth]{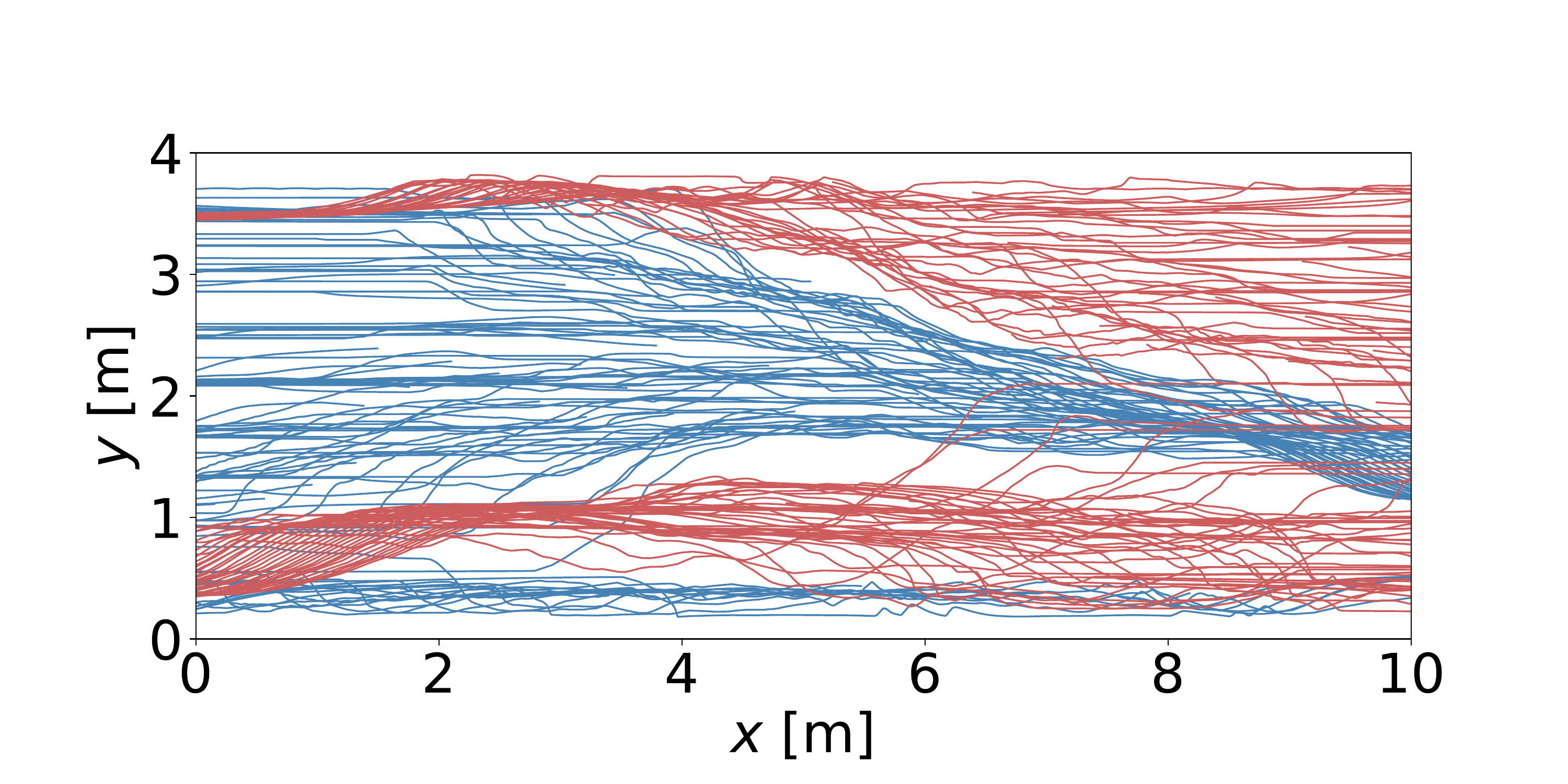}\label{fig:SimulationTraj}}
    \caption{
    The trajectories of pedestrians (agents) in the first \SI{50}{\second}.
    (a) Experiment.
    (b) Simulation.
    }
    \label{fig:ActualTraj}
\end{figure}

The first \SI{50}{\second} trajectories of agents in the simulation and pedestrians in the experiment are shown in Figure~\ref{fig:ActualTraj}.
Four lanes can be observed in both the experiment and the simulation.

\begin{figure}[H]
    \centering
    \subfigure[]{\includegraphics[width=0.4\linewidth]{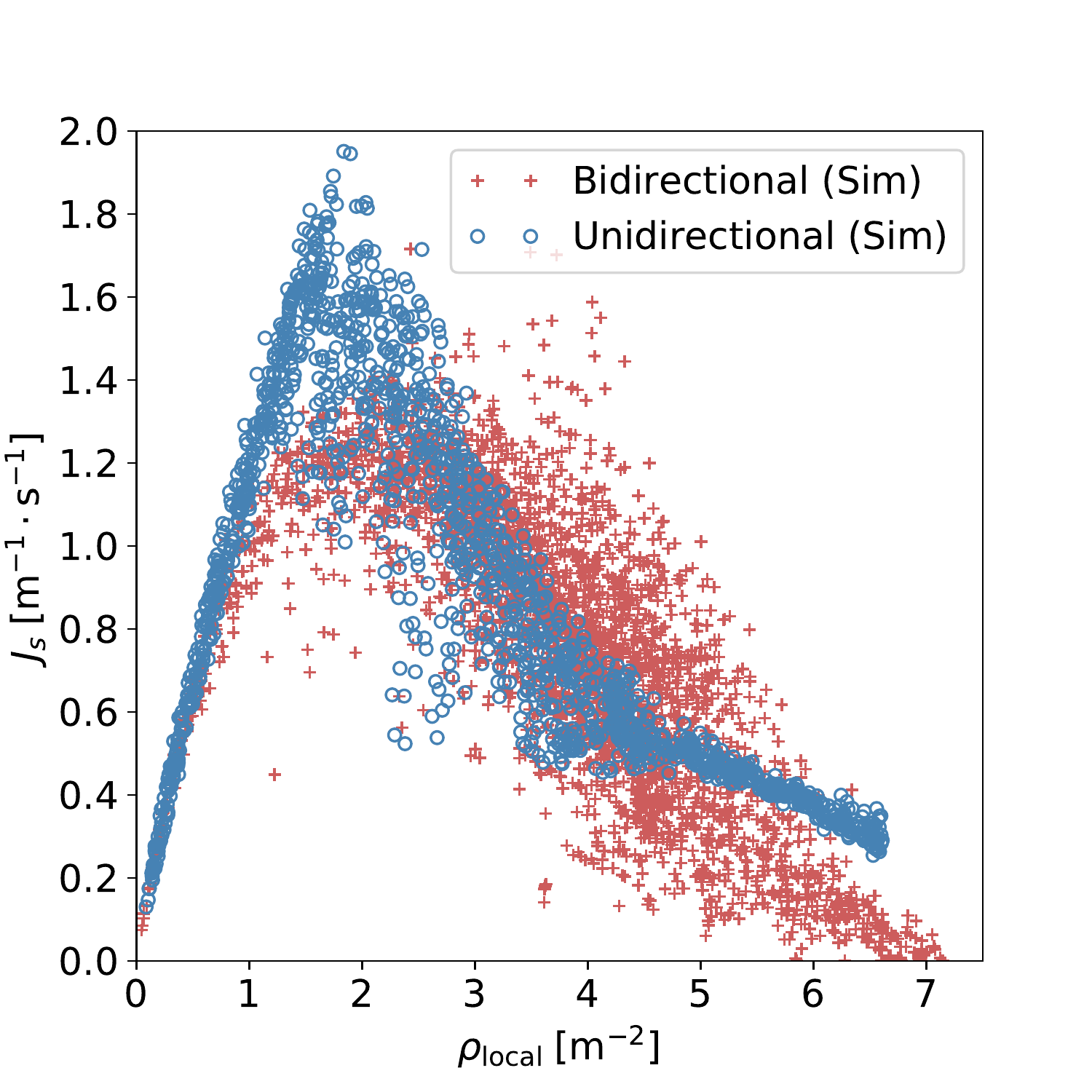}\label{fig:UniandBiFD}}
    \subfigure[]{\includegraphics[width=0.4\linewidth]{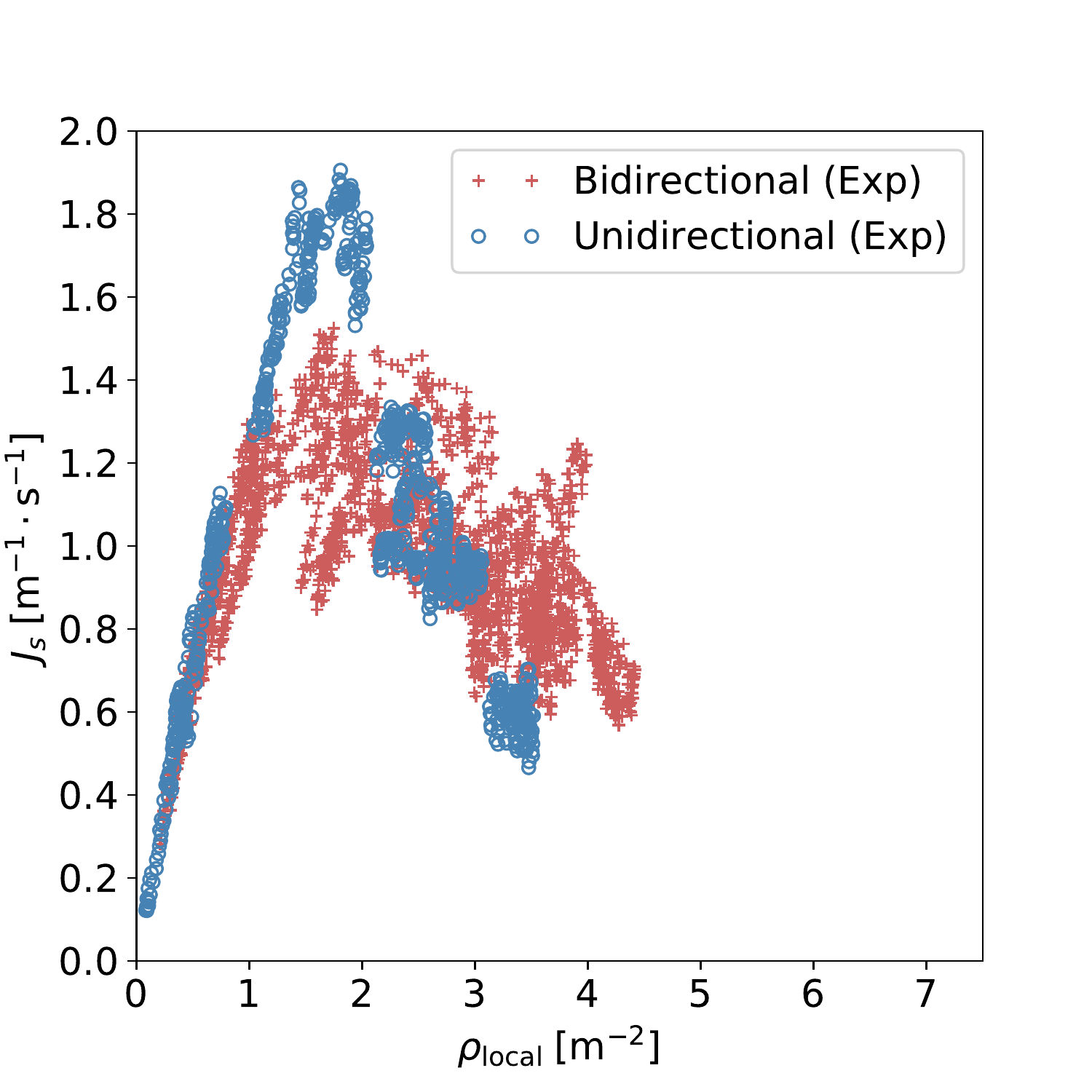}\label{fig:UniandBiFDExp}}
    \caption{The fundamental diagram of uni- and bidirectional flow.
    (a) The relation between the local density and the specific flow in the simulation.
    (b) The relation between the local density and the specific flow in the experiment.}
\end{figure}

The fundamental diagram of the unidirectional flow was then reproduced with the calibrated parameters listed in Table~\ref{tab:validatedPara}.
The FD of uni- and bidirectional flow obtained by simulations with the AVM are shown in Figure~\ref{fig:UniandBiFD}.
The specific flow reaches a peak with increasing local density $\rho_\text{local}$ in the unidirectional flow simulation, where a plateau is formed in the bidirectional flow simulation.
The difference is in line with the observation in the experiment~\cite{zhang2012ordering}.
See Figure~\ref{fig:UniandBiFDExp}.  
Note that the experimental data for higher densities than $\SI{4.5}{\meter^{-2}}$ cannot be reached.
Data from simulations and experiments show different scatter. 
The larger scatter at the congested regime indicates that real pedestrians steer more smoothly at high densities than the agents modeled by the AVM. 
One possible reason for this discrepancy is that the agents in the model always jostle to move further and are apparently unaware of the strategy of simply standing still.

\section{Conclusion}
\label{sec:conclusion}

A new velocity model is proposed to take into consideration anticipation of pedestrians.
For this, the process of anticipation is divided into three parts: perception of the actual situation, prediction of a future situation and selection of a strategy leading to an action.

First, the AVM is compared to the two other velocity-based models (the generalized collision-free velocity model (GCVM), and the collision-free speed model (CSM)) in binary interaction scenarios. 
Even in these simplified situations, the simulated trajectories of agents show that the AVM can reproduce the movement of pedestrians more realistically than the other two models.    

In a second step, these models are compared in the bidirectional flow scenario with periodic boundary conditions.
Simulations are classified as jamming state or moving state, according to the number of static, or blocked, agents in the simulation. 
Compared to the other two models, the critical density between the moving states and jamming states is shifted to higher values using the AVM.
This indicates that the AVM prevents imminent collisions better than the other models.

The influence of the parameters describing the effect from neighbors in the three models is studied in bidirectional flow scenarios.
Only for the AVM does an increase in the impact from neighbors in the direction of movement lead to an increase in the jamming probability. 
The opposite occurs when the CSM and the GCVM are used. 
The bidirectional flow simulation with periodic boundary conditions is then performed using the set of parameters that leads to the minimum jamming probability.
The jamming probability is significantly reduced in the simulation using the CSM and the AVM by adopting the new parameters but there is little change in the simulation using the GCVM.

After this, the quantity $\Phi$, describing the degree of order given by lanes, is adopted to analyze the formation of lanes quantitatively.
In line with experimental results of high-density situations, the AVM leads to the formation of lanes much faster than the CSM.
One possible reason for this difference here is that the moving state of the AVM can be attributed to the strategy of following, while the moving state of the CSM could be associated with agents pushing each other aside.

Finally, the AVM is validated using the fundamental diagram (FD).
After calibration based on the bidirectional FD, the FD for unidirectional flow is correctly reproduced by simulation using the AVM.
The difference between the FD of uni- and bidirectional flow is also well reproduced.
Moreover, the course of lane formation in time and the shape of the formed lanes in the simulations with the AVM are similar to those in the experiments.

Additional analyses using the AVM in the bidirectional flow simulation with periodic boundary conditions are conducted.
In \ref{sec:taAndAlpha}, the decisive factors leading to improved performance in the simulation using the AVM are studied.
The result shows that the prediction of the future situation and the strategy of following are both significant for reducing the jamming probability, and there is an optimal prediction time to realize the minimum jamming probability.
In \ref{sec:width}, simulations are performed in corridors with different widths.
When the corridor is wider than \SI{2}{\meter}, the effect of the corridor width on the jamming probability is insignificant.
In addition, the heterogeneity of agents' free speed is studied in \ref{sec:freeSpeed}, where the jamming probability decreases with increasing heterogeneity of the free speed of agents.

\section*{Acknowledgments}
The authors are grateful to the HGF for providing funding within the knowledge transfer project under Grant No. WT-0105.
Qiancheng Xu wishes to express his thanks for the funding support received from the China Scholarship Council (Grant No. 201706060186).
We would also like to thank Antoine Tordeux for his very useful comments and consultations.

\appendix
\section{Decisive factors in the anticipation velocity model}
\label{sec:taAndAlpha}
Using the AVM reduces the jamming probability and favors the formation of lanes in bidirectional flow simulations.
The possible causes for the improvement are the prediction time $t^\text{a}$, and the dynamic $\alpha_{i,j}$ reflecting pedestrians' preference to follow others moving in the same direction.
To identify the decisive factor of this improvement, simulations are performed in periodic boundary conditions with dynamic $\alpha_{i,j}$ and constant $\alpha_{i,j}$ ($\alpha_{i,j}=k$), respectively.
For each case, different values of $t^\text{a}$ (0, 0.2, 0.5, 1.0, 1.5, and \SI{2}{\second}) are adopted.
The simulation scenario is the corridor shown in Figure~\ref{fig:geo}.
The free speeds of agents are normally distributed $N\sim(1.55,0.18^2)~\SI{}{\meter\per\second}$.
The number of agents in each simulation is 140 ($\rho\approx\SI{1.35}{\per\square\meter}$).
This corresponds to the global density where the transition from moving states ($P_\text{jam}=0$) to jamming states ($P_\text{jam}=1$) occurs in the simulation using the AVM.
The other parameters of the AVM are shown in Table~\ref{tab:parameterCompare}.

\begin{figure}[H]
    \centering
    \subfigure[]{\includegraphics[width=0.45\linewidth]{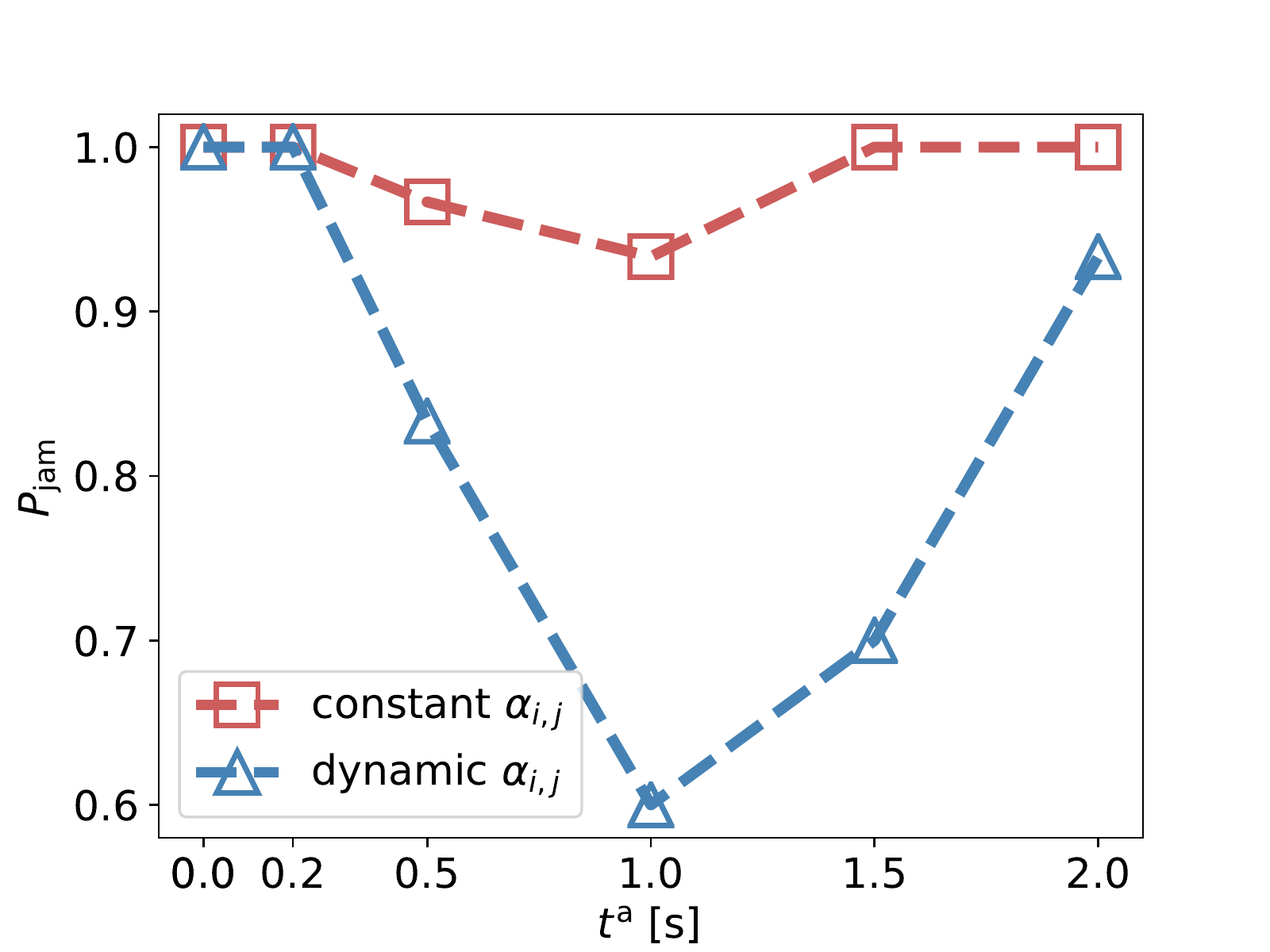}\label{fig:JamPTa}}
    \subfigure[]{\includegraphics[width=0.45\linewidth]{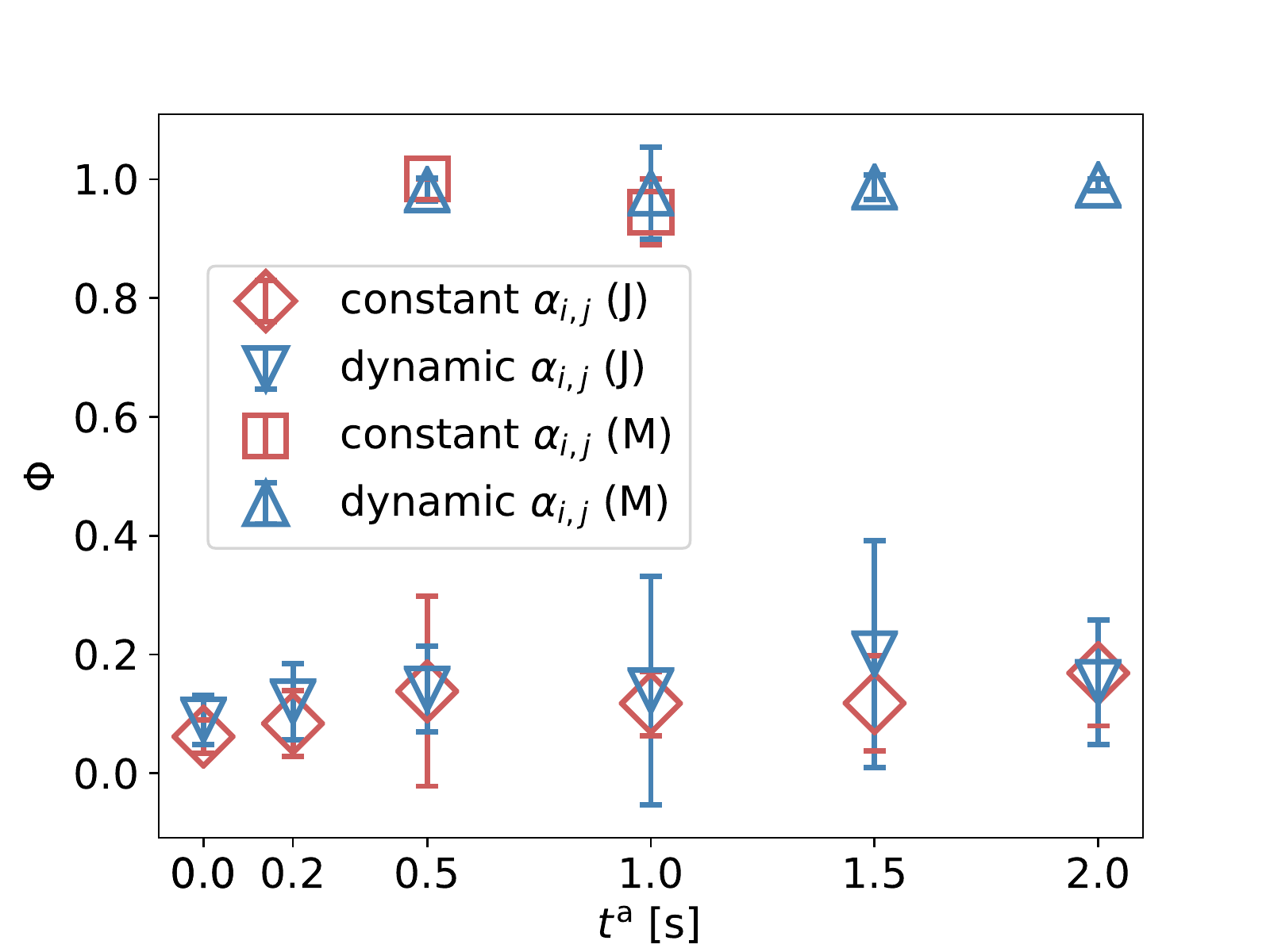}\label{fig:OrderTa}}
    \caption{
    (a): The relationship between the jamming probability $P_\text{jam}$ and the prediction time $t^\text{a}$ for dynamic and constant $\alpha_{i,j}$.
    (b): Mean value and standard deviation of $\Phi$ in the simulation with jamming (J) and moving (M) states for each value of $t^\text{a}$ and each $\alpha_{i,j}$ (dynamic and constant).
    }
\end{figure}

The relationship between the jamming probability $P_\text{jam}$ and the prediction time $t^\text{a}$ for dynamic and constant $\alpha_{i,j}$ is shown in Figure~\ref{fig:JamPTa}.
The same tendency of $P_\text{jam}$ is observed in the simulation with dynamic and constant $\alpha_{i,j}$.
As $t^\text{a}$ increases, $P_\text{jam}$ decreases until $t^\text{a}$ reaches a specific value, then it increases.
Moreover, the effect of the prediction time $t^\text{a}$ on the jamming probability $P_\text{jam}$ is more significant in the simulation with dynamic $\alpha_{i,j}$ than with constant $\alpha_{i,j}$.

The mean value and standard deviation of $\Phi$ in the simulation with jamming and moving states are calculated for each $t^\text{a}$ and each $\alpha_{i,j}$ (dynamic and constant).
See Figure~\ref{fig:OrderTa}.
The values of $\Phi$ are close to 1 in the moving state and less than 0.3 in the jamming state.
Neither the prediction time $t^\text{a}$ nor the coefficient $\alpha_{i,j}$ affects the formation of lanes in the simulation with the moving state.

In conclusion, the prediction time $t^\text{a}$ and the dynamic $\alpha_{i,j}$ both contribute to reducing the jamming probability in the bidirectional flow simulation.
Moreover, a longer prediction time does not mean a lower jamming probability.
An appropriate prediction time $t^\text{a}$ can reduce the jamming probability. 
A similar conclusion also drawn from~\cite{suma2012anticipation} is that there is an optimal strength of anticipation to realize the smoothest counterflow.

\section{Width of the corridor}
\label{sec:width}
A transition from moving states ($P_\text{jam}=0$) to jamming states ($P_\text{jam}=1$) occurs with an increase of $\rho_{\text{global}}$ (the global density of agents) in the bidirectional flow simulations.
In this appendix, the effect of the corridor width on this transition is studied.
Six corridors with different widths (1, 2, 3, 4, 5, and \SI{6}{\meter}) are simulated.
Apart from the width, these corridors are identical to the corridor shown in Figure~\ref{fig:geo}.
For each corridor, simulations are performed in periodic boundary conditions with different values of $\rho_{\text{global}}$ (0.31, 0.62, 0.92, 1,23, 1.54, or \SI{1.85}{\per\square\meter}).
The simulation is performed with the AVM using the parameters in Table~\ref{tab:parameterCompare}.
The free speeds of agents are normally distributed $N\sim(1.55,0.18^2)~\SI{}{\meter\per\second}$.

\begin{figure}[H]
    \centering
    \subfigure[]{\includegraphics[width=0.45\linewidth]{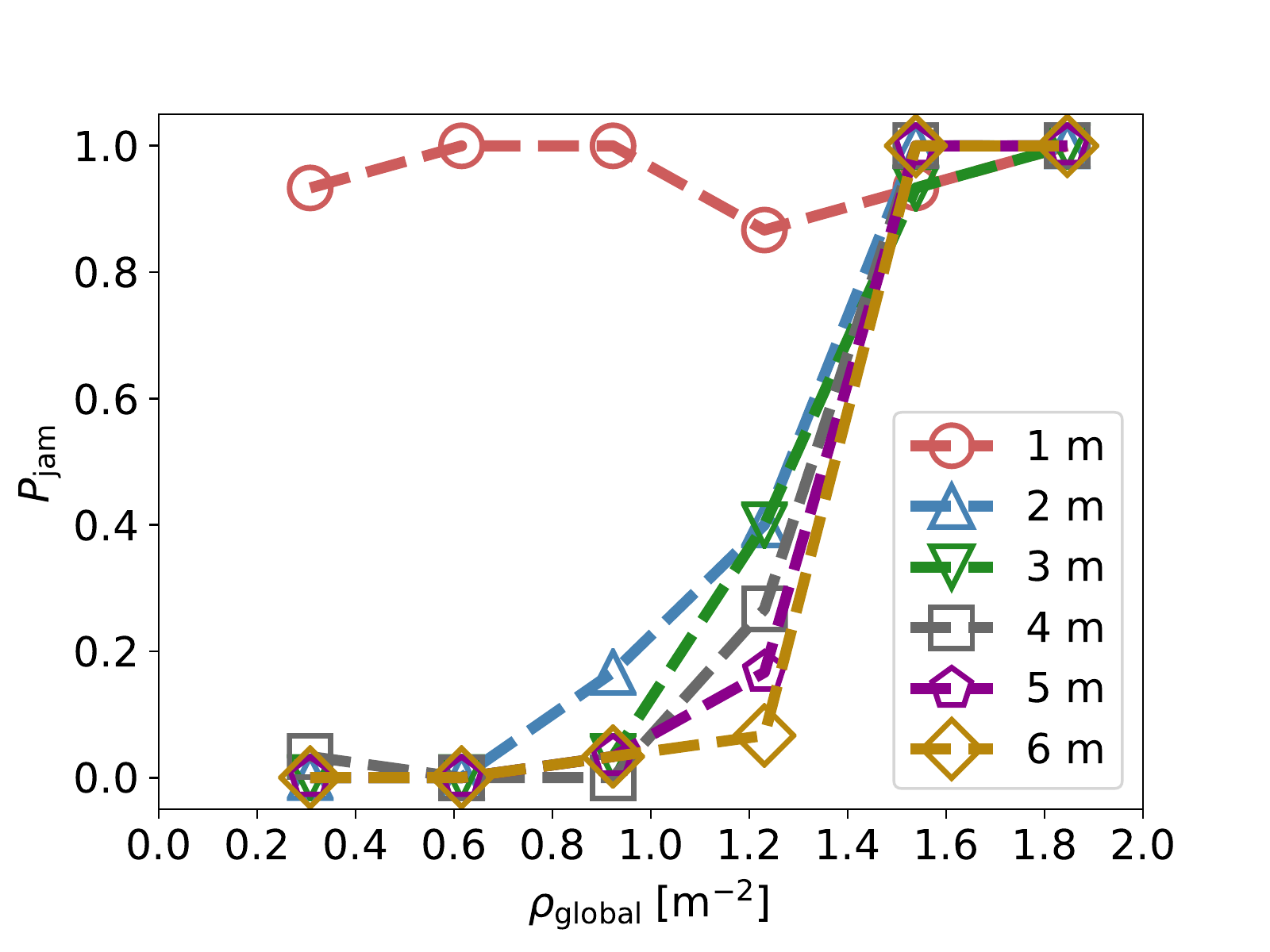}\label{fig:JamPWidthDensity}}
    \subfigure[]{\includegraphics[width=0.45\linewidth]{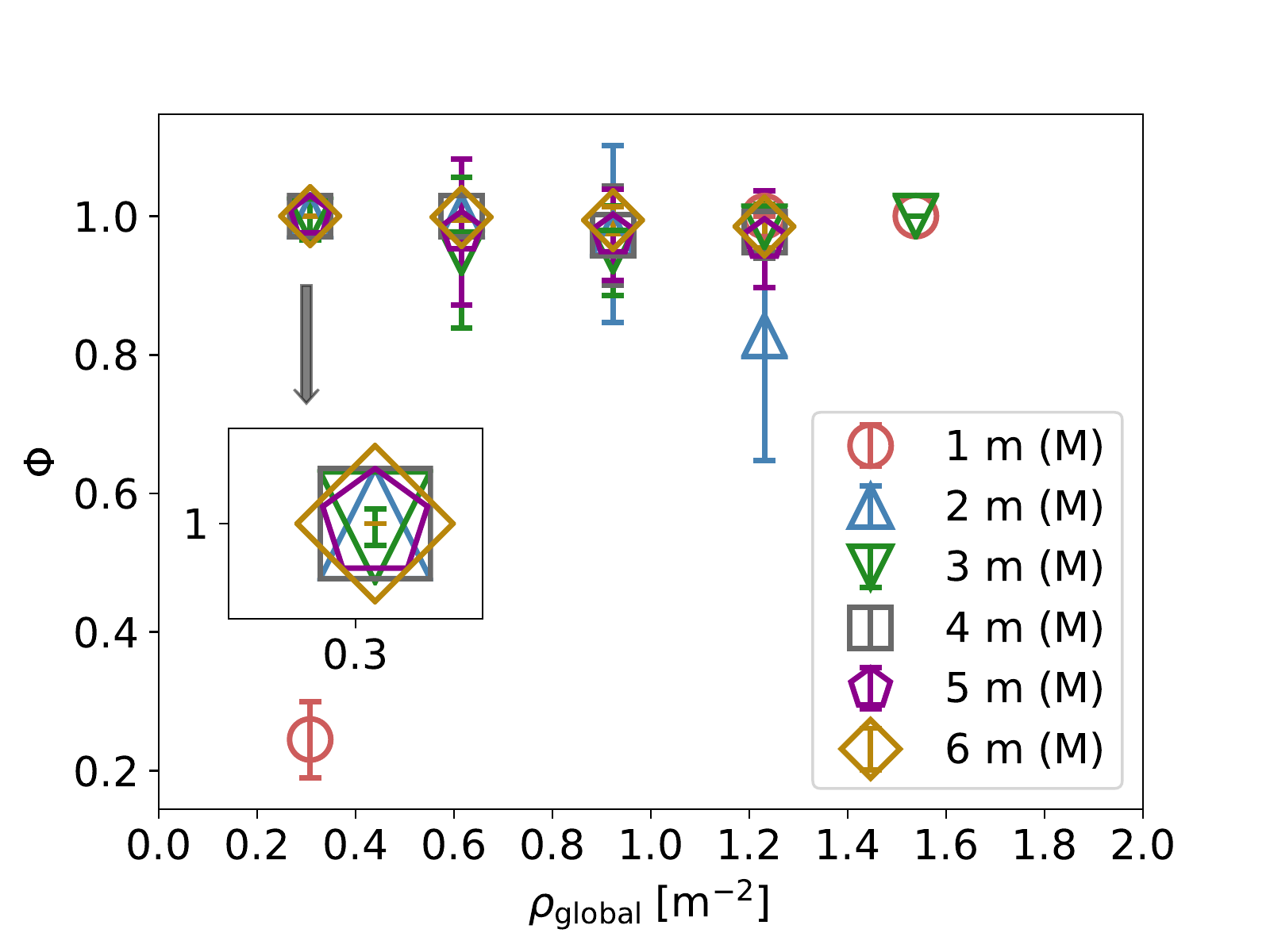}\label{fig:OrderWidth}}
    \caption{The numbers in the legend give the width of the corridor.
    (a): Relationship between the jamming probability $P_\text{jam}$ and the global density $\rho_{\text{global}}$ for corridors with different widths.
    (b): Mean value and standard deviation of $\Phi$ in the simulation with the moving state (M), for each corridor and each global density $\rho_{\text{global}}$.
    }
\end{figure}

The relationship between the jamming probability $P_\text{jam}$ and the global density $\rho_{\text{global}}$ for corridors with different widths is shown in Figure~\ref{fig:JamPWidthDensity}.
The transition from moving states to jamming states is observed in the simulation with all corridors except for the corridor of \SI{1}{\meter} width.
When the width of the corridor is \SI{1}{\meter}, the value of $P_\text{jam}$ is close to 1 even if the global density is very low. 
One possible reason for this is that the effect of walls prevents agents from using the full width of the corridor.

The mean value and standard deviation of $\Phi$ in the simulation with the moving state are calculated for each corridor and each value of $\rho_{\text{global}}$.
See Figure~\ref{fig:OrderWidth}.
When the width of the corridor is \SI{1}{\meter} and $\rho_{\text{global}}=\SI{0.31}{\per\square\meter}$, the value of $\Phi$ is lower than in other situations.
In addition to this, there is no significant difference between the value of $\Phi$ in other simulations with the moving state, which is always close to 1.

\section{Heterogeneity of the free speed of the agents}
\label{sec:freeSpeed}
To analyze the influence of the heterogeneity of agents, the free speed is chosen according to a normal distribution.
A larger standard deviation of the distribution corresponds to a higher heterogeneity in the free speed of agents.
In other sections, the normal distribution $N\sim(1.55,0.18^2)~\SI{}{\meter\per\second}$ obtained from the experiment in~\cite{zhang2012ordering} is used.
In this subsection, normal distributions with the same mean value (\SI{1.55}{\meter\per\second}) but different standard deviations (0, 0.09, 0.18, 0.36, or \SI{0.54}{\meter\per\second}), which is denoted by $\sigma$, are adopted in this analysis.
Simulations are performed with periodic boundary conditions in the corridor shown in Figure~\ref{fig:geo}.
The number of agents is 140 in each simulation ($\rho\approx\SI{1.35}{\per\square\meter}$).
It corresponds to the global density, where the transition from moving states ($P_\text{jam}=0$) to jamming states ($P_\text{jam}=1$) occurs in the simulation using the AVM.
Other parameters of AVM are shown in Table~\ref{tab:parameterCompare}.

\begin{figure}[H]
    \centering
    \subfigure[]{\includegraphics[width=0.45\linewidth]{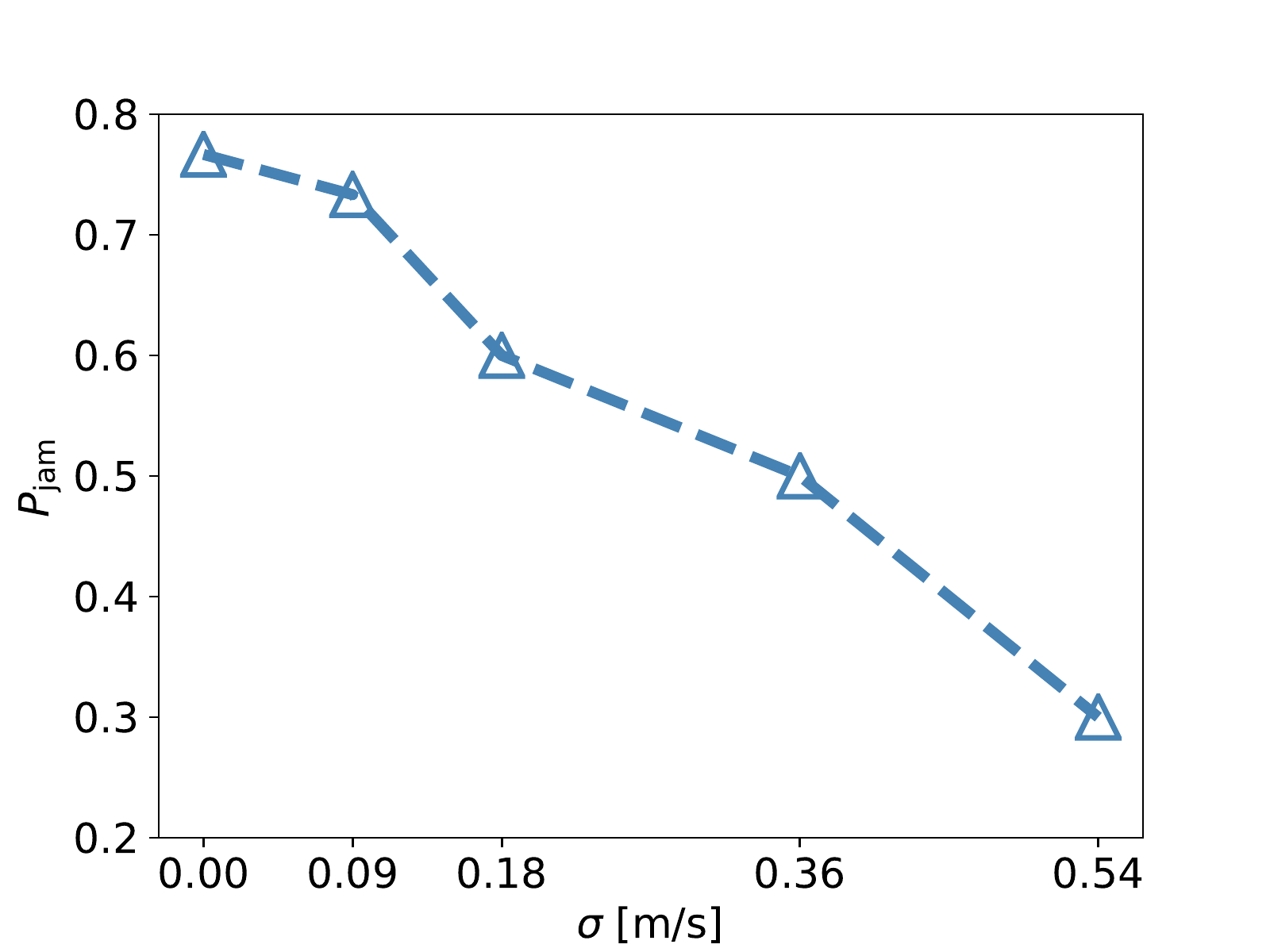}\label{fig:JamPV0}}
    \subfigure[]{\includegraphics[width=0.45\linewidth]{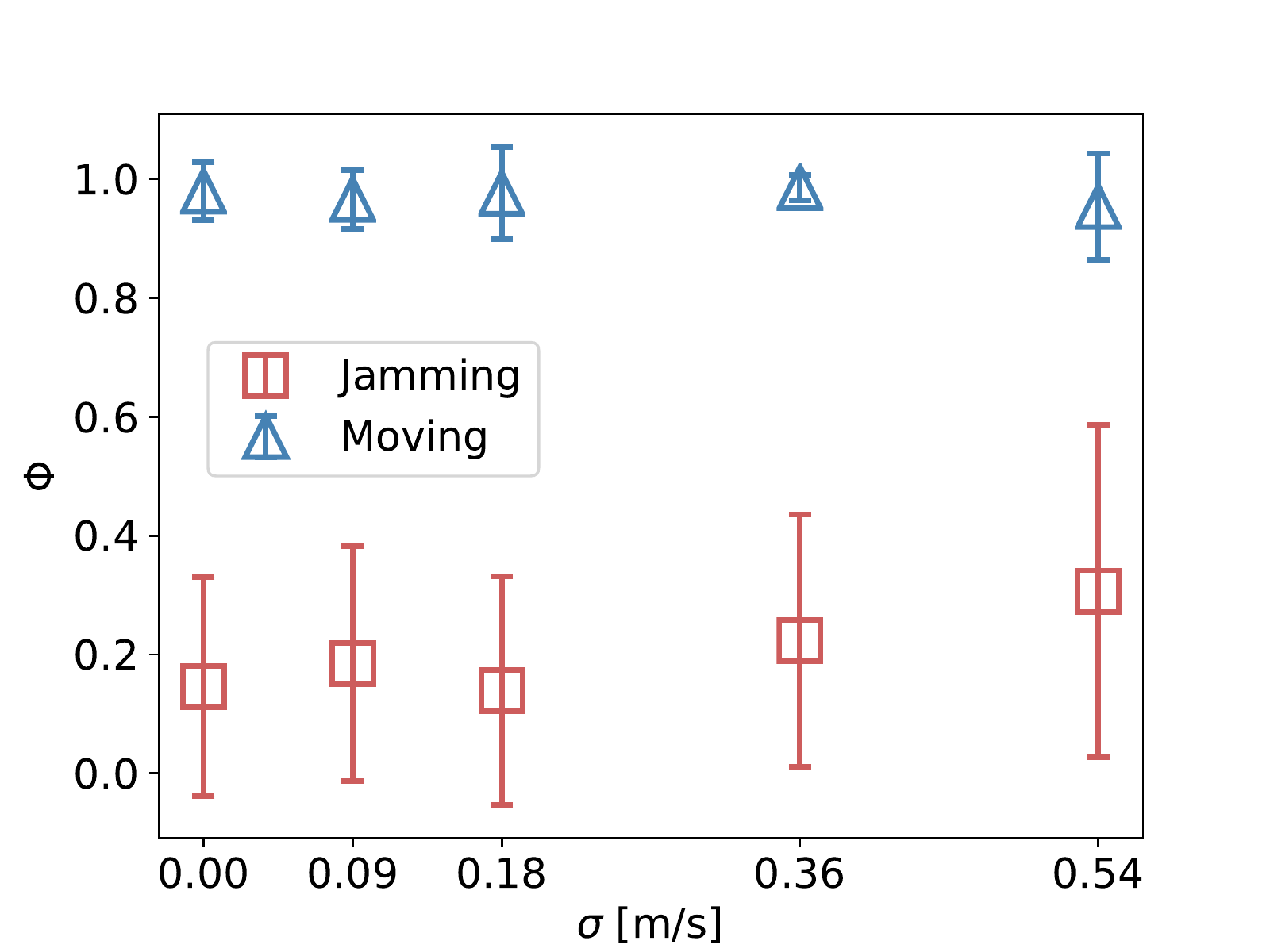}\label{fig:OrderV0}}
    \caption{
    (a): The relationship between the jamming probability $P_\text{jam}$ and the standard deviation $\sigma$.
    (b): The mean value and standard deviation of $\Phi$ in the simulation with jamming and moving states, for each value of $\sigma$.
    }
\end{figure}

The relationship between the jamming probability $P_\text{jam}$ and the standard deviation $\sigma$ is shown in Figure~\ref{fig:JamPV0}.
The value of $P_\text{jam}$ decreases as $\sigma$ increases.
The mean value and standard deviation of $\Phi$ in the simulation with jamming and moving states are calculated for each value of $\sigma$.
See Figure~\ref{fig:OrderV0}.
The value of $\Phi$ in the moving state is higher than in the jamming state.
Moreover, there is no significant difference between the value of $\Phi$ in the simulation with the moving state, which is always all close to 1.
In conclusion, with increasing heterogeneity of the free speed of agents, the jamming probability decreases, but the formation of lanes in the simulation with the moving state is not affected.

\bibliographystyle{elsarticle-num}
\bibliography{main}
\end{document}